\documentclass[%
twocolumn, superscriptaddress,
showpacs,preprintnumbers,
nofootinbib,
 amsmath,amssymb,
 aps,
pra,
]{revtex4}
\usepackage[all]{xy}
\usepackage{mathrsfs}
\usepackage{graphicx}
\usepackage{epstopdf}
\usepackage{dcolumn}
\usepackage{bm}
\usepackage{color} 
\usepackage{CJK}
\usepackage{ctable}  
\usepackage{hyperref}

\def\la{{\langle}}
\def\ra{{\rangle}}

\newcommand{\beq}{\begin{equation}}
\newcommand{\eeq}{\end{equation}}
\newcommand{\beqa}{\begin{eqnarray}}
\newcommand{\eeqa}{\end{eqnarray}}

\newcommand{\ua}{\uparrow}
\newcommand{\da}{\downarrow}

\begin{document}
\title{Fast phase gates with trapped ions}
\author{M. Palmero}
\email{mikel.palmero@ehu.eus}
\affiliation{Departamento de Qu\'{\i}mica F\'{\i}sica, UPV/EHU, Apdo.
644, 48080 Bilbao, Spain}
\author{S. Mart\' inez-Garaot}
\affiliation{Departamento de Qu\'{\i}mica F\'{\i}sica, UPV/EHU, Apdo.
644, 48080 Bilbao, Spain}
\author{D. Leibfried}
\affiliation{National Institute of Standards and Technology, 325 Broadway, Boulder, Colorado 80305, USA}
\author{D. J. Wineland}
\affiliation{National Institute of Standards and Technology, 325 Broadway, Boulder, Colorado 80305, USA}
\author{J. G. Muga}
\affiliation{Departamento de Qu\'{\i}mica F\'{\i}sica, UPV/EHU, Apdo.
644, 48080 Bilbao, Spain}
%
%
\begin{abstract}
We implement faster-than-adiabatic two-qubit phase gates using smooth state-dependent  forces. 
The forces are designed to leave no final motional excitation,  independently of the initial motional state in the harmonic, small-oscillations limit. They 
are simple, explicit functions of time and the desired logical phase of the gate, and are based on 
quadratic invariants of motion 
and Lewis-Riesenfeld phases of the normal modes. 
%
\end{abstract}
\maketitle
%
%
%
%
%
\section{Introduction}
%
%
%
Realizing the full potential of quantum information processing requires a sustained effort to 
achieve scalability, and to make basic dynamical or logical operations faster, more accurate and reliable under perturbations.   
Two-qubit gates are crucial building blocks in any scheme of universal quantum computing and 
have received much attention. 
An important step forward was the theoretical proposal of geometric gates with reduced sensitivity  to the vibrational quantum numbers
\cite{Sorensen1999,Solano1999,Milburn2000,Sorensen2000}, with the first experimental realization in \cite{Sackett}.
Soon after, Leibfried et al. \cite{Leibfried2003} 
demonstrated
a phase gate of the form 
\beqa
|\uparrow\uparrow\ra\to |\uparrow \uparrow\ra,\; |\downarrow\downarrow\ra\to|\downarrow\downarrow\ra,
\nonumber
\\ 
|\uparrow\downarrow\ra\to i |\uparrow\downarrow\ra,\; |\downarrow\uparrow\ra\to i |\downarrow\uparrow\ra,
\label{gate}
\eeqa
with two trapped ions of the same species subjected to state-dependent forces,
where each spin-up/down arrow represents an eigenstate of the $\sigma_z$-operator for one of the ion-qubits.    
Generalizations of this gate with the potential of reduced gate times were discussed by 
Garc\'\i a-Ripoll et al. \cite{Garcia-Ripoll2003,Garcia-Ripoll2005}, 
and in \cite{Staanum2004,Vitanov2014,Steane2014}.
The gate mechanism satisfies a number of 
desirable properties: it is insensitive to the initial motional state of the ions,  at least in the  small-oscillations regime
{where the motion is inside the Lamb-Dicke regime and the nonlinearities of the Coulomb coupling are negligible}; 
it depends on ``geometric'' properties of the dynamics (phase-space areas), which makes it resistant to 
certain errors;  it allows for close distances and thus strong interactions among the ions; and, 
finally, it may in principle be driven in short, faster-than-adiabatic times.      
The forces designed to make the ions return to their initial motional state in a rotating
frame of phase-space coordinates \cite{Milburn2000, Sorensen2000}, are different 
for different qubit state configurations, 
leading to qubit-state dependent motional trajectories that produce a differential  phase. 
Pulsed forces with abrupt kicks 
were designed \cite{Steane2014,Bentley2013,Taylor2016,Duan2004}, 
and also  forces with discontinuous  but finite derivatives \cite{Staanum2004,Garcia-Ripoll2005,Vitanov2014},
at boundary times of the entire operation, or between pulses,
but 
in practice 
smooth continuous forces with continuous derivatives are
desirable to minimize experimental errors. 

In this work, we revisit the phase gates and tackle the design of smooth forces as an inverse problem, 
via Lewis-Riesenfeld invariants \cite{Lewis1969}. For a predetermined operation time, the approach is 
applicable to arbitrary masses, and 
proportionalities among the spin-dependent forces.   
This provides a more general scope than previous proposals to achieve faster than
adiabatic operations.    
Hereafter, forces are assumed to be induced by off-resonant
lasers that do not change the internal states. However, the basic ideas should be applicable to M\o lmer and S\o rensen 
type gates that flip the qubit spins during gates as well\footnote{The M\o lmer \& S\o rensen gate \cite{Sorensen1999,Sorensen2000} can be mathematically described in the same language, 
replacing the eigenvectors of  $\sigma_z$, $|\uparrow\rangle$ and $|\downarrow\rangle$, by the 
eigenvectors of $\sigma_x$,  $|+\rangle$ and $|-\rangle$. This allows for an interchange 
of methods among the gate (\ref{gate2}) and the M\o lmer \& S\o rensen gate.}  \cite{Lee2005}.
Specifically we design forces to implement the operation    
\beqa
|\uparrow\uparrow\ra\to e^{i\phi(\uparrow\uparrow)}|\uparrow \uparrow\ra,\; 
|\downarrow\downarrow\ra\to e^{i\phi(\downarrow\downarrow)}|\downarrow\downarrow\ra,
\nonumber
\\ 
|\uparrow\downarrow\ra\to e^{i\phi(\uparrow\downarrow)}|\uparrow\downarrow\ra,\; 
|\downarrow\uparrow\ra\to e^{i\phi(\downarrow\uparrow)} |\downarrow\uparrow\ra,
\label{gate2}
\eeqa
such that $\Delta\phi\equiv \phi(\uparrow\downarrow)+\phi(\downarrow\uparrow)-\phi(\uparrow\uparrow)-\phi(\downarrow\downarrow) =\pm \pi$, where the qubits could be realized with two different species,  
which may have practical importance to scale up quantum information processing with 
trapped ions \cite{Tan2015}. Gates of the form (\ref{gate2}) are computationally equivalent, up to single-qubit $z-$rotations to the  standard phase gate ${\rm diag}[{1,1,1,-1}]$ written in the basis $\{{|\uparrow\uparrow\ra, |\uparrow\downarrow\ra,|\downarrow\uparrow\ra,|\downarrow\downarrow\ra}\}$  \cite{Sasura2003}.

Our analysis demonstrates that invariant-based inverse Hamiltonian design  is not 
limited to population control and may be adjusted for phase control as well. 
It was known that the phase of a given mode of the invariant (a time-dependent eigenstate 
of the invariant which is also a solution of the time-dependent  Schr\"odinger equation) could be controlled \cite{Chen2010}, 
but the fact that ``global phases'', for a given internal state configuration,   of arbitrary motional states can be controlled as well 
in a simple way had been
overlooked. This is interesting for applying shortcuts to adiabaticity \cite{Torrontegui2013} in quantum information processing.
In particular, we will derive ready-to-use, explicit expressions for the state-dependent forces, and may benefit from the design freedom
offered by the invariant-based method
to satisfy further optimization criteria.         

To evaluate the actual performance of the phase gate at short times we have to compute fidelities,
excitation energies and/or their scaling behavior,   
according to the dynamics implied by the Hamiltonian including the anharmonicity of the Coulomb repulsion. 
This is important, as inversion protocols that work near perfectly in the   small-oscillations
regime,  fail for the large amplitudes of ion motion that occur in fast gates, and only a rough estimate of the 
domain of validity could be found in \cite{Garcia-Ripoll2005}. Here, we have numerically checked the validity of the phase gate up to gate
times less than one oscillation period without assuming the approximations used in the small-amplitude regime.  
An additional perturbing effect  with respect to an idealized limit of homogeneous spin-dependent forces 
is the position dependence of the forces induced by optical beams. 
This may be serious at the large motional amplitudes required for short gate times, when the ion motion amplitude becomes comparable 
to the optical wavelength as we illustrate with numerical examples. 


The analytical theory for small oscillations is worked out in Secs. {\ref{I}}, {\ref{II}}, and {\ref{III}}.
Then, we consider in Sec. \ref{IV}
two ions of the same species, which implies some simplifications, assuming  equal forces on both ions if they are in
the same internal state, and opposite forces with equal magnitude when the ions are in different internal states
(more general forces are treated in Appendix \ref{AppendixB}).  
We also consider a more complete Hamiltonian including the anharmonicity of the Coulomb force and the spatial dependence of the light fields  
to find numerically the deviations with respect to ideal results within the small oscillations approximation. 
Finally, in Sec. \ref{V} we study phase gates between ions of different species.  
The appendices present:  generalizations
of the results for arbitrary proportionalities between the state-dependent forces, 
alternative useful  expressions for the phases, 
an analysis to determine the worst possible fidelities, the calculation of the width of the position of one ion 
in the two-ion ground state, and alternative inversion protocols.    

%
%
%
%
%
\section{The model\label{I}}
Consider two ions of charge $e$, masses $m_1$, $m_2$, and coordinates $x_1$, $x_2$,   
trapped within the same, radially-tight, effectively one-dimensional (1D) trap. 
We assume the position $x_1$ of  ``ion 1'' to fulfill $x_1<x_2$ at all times due to Coulomb repulsion, with $x_2$ the position of ``ion 2''. Qubits 
may be encoded for each ion in two internal levels corresponding to ``spin up'' ($|\uparrow\rangle$) eigenstate of $\sigma_z$ with eigenvalue $\sigma_i^z=1$, and ``spin down'' eigenstate $(|\downarrow\ra)$ with eigenvalue $\sigma_i^z=-1$, 
$i=1,2$.  
Off-resonant lasers induce state dependent forces that are assumed first to be homogeneous over the extent of 
the motional state (Lamb-Dicke approximation).
Later in the text we shall analyze the effect of more realistic position-dependent light fields when the Lamb-Dicke condition is not satisfied. 
For a given spin configuration, $\uparrow\uparrow$, $\downarrow\downarrow$, $\uparrow\downarrow$, or $\downarrow\uparrow$, the Hamiltonian can be written as
\beqa
\label{Hamiltonian}
{\cal H} &=& \frac{p_1^2}{2m_1} + \frac{1}{2} u_0x_1^2 + F_1(t;\sigma_1^z)x_1
\nonumber\\
&+& \frac{p_2^2}{2m_2} + \frac{1}{2} u_0x_2^2 + F_2(t;\sigma_2^z)x_2
\nonumber\\
&+& \frac{C_c}{x_2-x_1}-E_0,
\eeqa
where $C_c=\frac{e^2}{4\pi\epsilon_0}$, $u_0=m_1\omega_1^2=m_2\omega_2^2$, and $\epsilon_0$ is the vacuum permittivity. 
A constant $E_0$ is added for convenience so that the minimum of
\beq 
{\cal V}= \frac{1}{2} u_0x_1^2 +  \frac{1}{2} u_0x_2^2 + \frac{C_c}{x_2-x_1} -E_0 
\eeq
is at zero energy  when $x_1$ and $x_2$ assume their equilibrium positions.  
The laser induced, state-dependent forces may be independent for different  ions as they may 
be implemented by different lasers on different transitions. For equal mass ions, the same lasers and equal and opposite forces on the qubit eigenstates, 
 they may simplify to $F_i=\sigma_i^zF(t)$. 
In principle, the proportionality between the force for the up and the down state could be different, 
but, as shown in  Appendix \ref{AppendixB},
the forces for a general proportionality can be related by a simple scaling to the ones found for the symmetric case $F_i=\sigma_i^zF(t)$.

We can determine normal modes for the zeroth order Hamiltonian  
\beq
\label{zeroth}
{\cal H}_0=\frac{p_1^2}{2m_1} + \frac{p_2^2}{2m_2} +{\cal V}. 
\eeq
The equilibrium positions of both ions under the potential ${\cal V}$ are 
\beq
x_1^{(0)}=-\sqrt[3]{\frac{C_c}{4u_0}}, \quad x_2^{(0)}=\sqrt[3]{\frac{C_c}{4u_0}},
\eeq
with equilibrium distance  $x_0=x_2^{(0)}-x_1^{(0)}$, which yields $E_0=3 u_0 x_0^2/4$. 

Diagonalizing the mass scaled curvature matrix $V_{ij}=\frac{1}{\sqrt{m_im_j}}\frac{\partial^2{\cal V}}{\partial x_i\partial x_j}\big|_{\{x_i,x_j\}=\{x_i^{(0)},x_j^{(0)}\}}$, 
 that describes the restoring forces for small oscillations around the equilibrium positions, we get the eigenvalues
\beq
\lambda_\pm =\omega_1^2\left[1+\frac{1}{\mu}\pm\sqrt{1-\frac{1}{\mu}+\frac{1}{\mu ^2}}\right],
\eeq
where $\omega_1=(u_0/m_1)^{1/2}$ and $\mu = m_2/m_1$, with $\mu\geq 1$. 
The normal-mode angular frequencies  are  
\beq
\Omega_\pm=\sqrt{\lambda_\pm}, 
\eeq
and the orthonormal eigenvectors take the form  
$v_\pm=\left(\begin{array}{c}a_\pm\\b_\pm\end{array}\right)$, where 
\beqa
a_\pm&=&\left[\frac{1}{1+\left(1-\frac{1}{\mu}\mp\sqrt{1-\frac{1}{\mu}+\frac{1}{\mu ^2}}\right)^2\mu}\right]^{1/2},
\nonumber\\
b_\pm&=&\left(1-\frac{1}{\mu}\mp\sqrt{1-\frac{1}{\mu}+\frac{1}{\mu ^2}}\right)\sqrt{\mu}a_\pm,
\eeqa
fulfill 
\beqa
a_\pm^2+b_\pm^2&=&1,
\nonumber\\
a_+a_-+b_+b_-&=&0,
\nonumber\\
a_+b_--a_-b_+&=&1.
\label{abrela}
\eeqa
The mass-weighted, normal-mode coordinates are
\beqa
\label{NMcoordinates}
{\sf x}_+ &=& a_+\sqrt{m_1}(x_1-x_1^{(0)}) + b_+\sqrt{\mu m_1} (x_2-x_2^{(0)}),
\nonumber\\
{\sf x}_- &=& a_-\sqrt{m_1}(x_1-x_1^{(0)}) + b_-\sqrt{\mu m_1} (x_2-x_2^{(0)}),
\eeqa
and the inverse transformation to the original position coordinates is  
\beqa
x_1 &=& \frac{1}{\sqrt{m_1}}(b_-{\sf x}_+-b_+{\sf x}_-)-\frac{x_0}{2},
\nonumber\\
x_2 &=& \frac{1}{\sqrt{\mu m_1}}(-a_-{\sf x}_++a_+{\sf x}_-)+\frac{x_0}{2}.
\label{x1x2}
\eeqa
Finally, the Hamiltonian (\ref{Hamiltonian}), neglecting higher-order anharmonic 
terms, and using conjugate ``momenta'' ${\sf p}_\pm=-i\hbar\partial/\partial{\sf x}_\pm$,\footnote{The dimensions of the mass weighted 
coordinates are length times square root of mass,  $m kg^{1/2}$, while the dimensions of the conjugate momenta are $kg^{1/2}m/s$.} takes the form
\beq
\label{HamiltonianNM}
{\cal H}=H_{NM}+\tilde{f}(t),
\eeq
where
\beqa
\label{hanm}
H_{NM} &=& H_++H_-,
\nonumber\\
H_\pm&=&\frac{{\sf p}_\pm^2}{2} + \frac{1}{2}\Omega_\pm^2 {\sf x}_\pm^2  -{f_\pm}{\sf x}_\pm,
\nonumber\\
\tilde{f}&=&\frac{x_0}{2}(F_2-F_1), 
\nonumber\\
f_\pm (t)&=&\mp \frac{F_1b_\mp}{\sqrt{m_1}}\pm \frac{F_2a_\mp}{\sqrt{\mu m_1}}.
\eeqa
The function $\tilde{f}$ depends on time and on the internal states.
By restricting the calculation to 
a given spin configuration, the dynamics may be worked out in terms of $H_{NM}$ alone, 
$i\hbar \partial{\psi_{NM}}/{\partial t}=H_{NM}\psi_{NM}$, and   
the wave function that evolves with ${\cal H}$ in Eq. (\ref{HamiltonianNM}) is   
$e^{(-i/\hbar)\int_0^{t} dt'\tilde{f}}Ê\psi_{NM}$. 
Purely time-dependent terms in the Hamiltonian are usually ignored as they imply global phases. 
 In the phase-gate scenario, however, they are not really global, since they depend on the spin  configuration. As the spin configuration may be changed after applying the phase gate, 
e.g. by resonant interactions, they may lead to observable interference effects and  in general cannot be ignored. 
However, in the particular gate operation studied later the extra phase vanishes at the final time $t_f$, so we shall   
focus on the dynamics and phases generated by    
the Hamiltonian $H_{NM}$, which   
represents  
two independent  forced harmonic oscillators with constant frequencies. 
We can now apply Lewis-Riesenfeld theory \cite{Lewis1969} in an inverse way \cite{Torrontegui2013}: 
The desired dynamics are designed first, and from the corresponding invariant    
the time-dependent functions in the Hamiltonian are inferred \cite{Torrontegui2011}. Note   
that in the inverse problem the oscillators are ``coupled'', as only one physical 
set of forces that will act on both normal modes of the uncoupled system 
must  be designed \cite{Palmero2014}.     
\section{One mode\label{II}}
In this section we consider just one mode and drop the subscripts $\pm$ to make the treatment applicable to both modes. 
The goal is to find expressions for the corresponding invariants, dynamics, and phases. 
The Hamiltonian describing a harmonic oscillator with mass-weighted position and momentum
is written as  
\beqa
H&=&H_0+V,
\\
H_0&=&\frac{p^2}{2}+\frac{1}{2}\Omega^2 x^2,
\\
V&=&-f(t)x.
\eeqa
Following the work of Lewis and Riesenfeld \cite{Lewis1969}, it is possible to find a dynamical invariant of $H$
solving the equation
\beq
\label{inva}
\frac{d I}{d t}\equiv\frac{\partial I}{\partial t}+\frac{1}{i\hbar}[I,H]=0.
\eeq
For a moving harmonic oscillator, a simple way to find an invariant is to assume a quadratic (in position and momentum)
ansatz with parameters that may be determined by inserting the ansatz in Eq. \eqref{inva}. 
This leads to the invariant
\beq
\label{invarianta}
I(t)= \frac{1}{2}({ p}-\dot{y})^2 + \frac{1}{2} \Omega^2 ({x}-y)^2,
\eeq
where the dot means ``time derivative'', and  the function $y(t)$ 
must satisfy the differential (Newton) equation
\beq
\label{auxiliarya}
\ddot{y}+\Omega^2 y=f,
\eeq
so it can be interpreted as a ``classical trajectory'' (with dimensions $kg^{1/2} m$)
in the forced harmonic potential \cite{Torrontegui2011}.
 
This invariant is Hermitian, and has a complete set of eigenstates. Solving
\beq\label{Ieigeneq}
I(t) \psi_n(t)=\lambda_n\psi_n(t),
\eeq
we get the time-independent eigenvalues
\beq\label{Ieigenval}
\lambda_n=\hbar\Omega\left(\frac{1}{2}+n\right),
\eeq
and the time-dependent eigenvectors
\beq
\label{psina}
\psi_n({x}, t)=e^{\frac{i}{\hbar}\dot{y}{{x}}} \phi_n\left({x}-y \right),
\eeq
where $\phi_n(x)$ is the $n$th eigenvector of the stationary oscillator,
\beq
\label{psistat}
\phi_n(x)=\frac{1}{\sqrt{2^n n!}}\left(\frac{\Omega}{\pi\hbar}\right)^{1/4} e^{\frac{-\Omega x^2}{2\hbar}} H_n\left(\sqrt{\frac{\Omega}{\hbar}} x\right), 
\eeq
and the $H_n$ are Hermite polynomials.  
The Lewis-Riesenfeld phases $\theta_n$ must satisfy  
\beq\label{LRpha}
\hbar\frac{d\theta_n}{dt}=\left\langle \psi_n\left|i\hbar\frac{\partial}{\partial t}-H\right|\psi_n\right\rangle,
\eeq
so that the wavefunction (\ref{solua}) is indeed a solution of the time-dependent Schr\"odinger equation. Using Eq. (\ref{psina}), they are given by 
\beqa
\theta_n(t) &=&-\frac{1}{\hbar} \int _0^t dt'(\lambda_n+\dot{y}^2/2-\Omega^2y^2/2)
\nonumber\\
&=& -(n+1/2)\Omega t-G(t),
\label{phasen}
\eeqa
where
\beq
G(t)=\frac{1}{2\hbar} \int _0^t dt'(\dot{y}^2-\Omega^2y^2).
\eeq
Finally, the solution of the Schr\"odinger equation for the Hamiltonian $H$ can be stated in terms of the eigenstate and Lewis-Riesenfeld phases of the invariant as
\beq\label{solua}
\psi({x},t)=\sum_n c_n e^{i\theta_n(t)}\psi_n({x},t).
\eeq
Hereafter we consider that $f$ is such that there are particular solutions $y=\alpha$ of Eq. (\ref{auxiliarya}) that satisfy at the boundary times $t_b=0,t_f$ the 
boundary conditions
\beq
\label{conds1a}
\alpha(t_b)=\dot{\alpha}(t_b)=0.  
\eeq
They guarantee that   
all states  $\Psi_n(x,t)=e^{i\theta_n^{(\pm)}(t)}\psi_n({x},t)$ end up at the original positions and at rest,
\beq
\Psi_n(x,t_f)=e^{i\theta_n(t_f)} \phi_n\left({x}\right). 
\eeq
In other words, each initial eigenstate of the Hamiltonian is driven along a path that returns to the initial state
with an added path-dependent phase.     
Moreover, we assume that the force vanishes at the boundary times $t_b=0,t_f$, $f(t_b)=0$, and, therefore, from Eq. (\ref{auxiliarya}),  
\beq
\label{conds2a}
\ddot{\alpha}(0)=\ddot{\alpha}(t_f)=0.
\eeq
Integrating by parts and using Eq. (\ref{auxiliarya}) as well as the boundary conditions $\alpha(t_b)=0$, the phase 
factor common to all $n$ takes the form 
\beqa
\label{phasea}
\phi(t_f)&=&-G(t_f)=\frac{1}{2\hbar}\int_0^{t_f}\!\! dt' f\alpha.
\eeqa
To determine the stability of the results with respect to a systematic perturbation let us assume that the force 
is subjected to a homogeneous, small constant offset $\delta_f$. Substituting $f\to f+\delta_f$, in first order, the phase (\ref{phasea}) is perturbed as  
\beq
\label{Delphasea}
\Delta\phi(t_f)=\frac{\delta_f}{2\hbar}\int_0^{t_f}\!\! dt' \alpha.
\eeq
Note that this could vanish if the zeroth order trajectory $\alpha$ nullifies the integral, as it happens  for the symmetrical 
functions used in this paper.  

As the phases $\theta_n(t_f)$ in Eq. (\ref{phasen}) have an extra  $n$-dependent term, 
an arbitrary motional state $\psi(t)$ that superposes different $n$-components does not generally return   
to the same initial projective ray. To remedy this it is useful to consider a rotating frame, i.e., 
we define $\psi_I(t)=e^{iH_0 t/\hbar}\psi(t)$, 
so that 
\beq
\psi_I(t_f)=e^{-iG(t_f)}\psi_I(0),
\eeq
with total phase $-G(t_f)$ for an arbitrary motional state. 
To decompose this phase into dynamical and geometric phases, we first note that 
\beq
i\hbar\frac{\partial \psi_I}{\partial t}=V_I\psi_I,
\eeq
where $V_I=-f e^{iH_0t/\hbar} x e^{-iH_0t/\hbar}$. 
The dynamical phase is  
\beqa
\phi_d&=&-\frac{1}{\hbar}\int_0^{t_f} dt \la\psi_I(t)|V_I(t)|\psi_I(t)\ra
\nonumber\\
&=& -\frac{1}{\hbar}\int_0^{t_f} dt \la\psi(t)|V(t)|\psi(t)\ra
\nonumber\\
&=&\frac{1}{\hbar}\int_0^{t_f} dt f(t)\la x(t)\ra.
\label{phased}
\eeqa
The expectation value of $x$ corresponds to a classical trajectory, i.e., to a solution of Eq. (\ref{auxiliarya}), but not necessarily 
the one corresponding to $\alpha$. To describe a general trajectory   
it is useful to define dimensionless positions and momenta as 
\beq
Y=\sqrt{\frac{\Omega}{2\hbar}}\,y,\;\;\;
P=\sqrt{\frac{1}{2\hbar \Omega}}\, p,
\eeq
(similarly for other coordinates such as $x$ or  $\alpha$) as well as complex-plane combinations $z=Y+iP$. 

The general solution of the position and momentum of a classical particle, or the corresponding expectation values 
for any quantum state 
is compactly given in complex form as
\beqa
z_g(t)&=&e^{-i\Omega t}\left\{z_g(0)+\frac{i}{\sqrt{2\hbar\Omega}}\int_0^t d\tau e^{i\Omega\tau}  f  \right\}
\nonumber\\
&=&\tilde{z}+z_0,
\label{gene}
\eeqa
where
\beqa
\tilde{z}&\equiv& e^{-it\Omega} z_g(0),
\\
z_0&\equiv&\sqrt{\frac{\Omega}{2\hbar}}y_0+i \sqrt{\frac{1}{2\Omega\hbar}}\dot{y_0},
\eeqa
and $y_0$ is a particular solution satisfying $y_0(0)=\dot{y_0}(0)=0$. 
For an $f$ such that $y_0(t)=\alpha(t)$, and thus $z_0=z_\alpha$, 
the boundary conditions at $t_f$ are satisfied as well in the particular solution, see  Eq. (\ref{conds1a}).  
By separating into real and imaginary parts, it can be seen that 
\beq
\Re{\rm e} (\tilde{z}) \frac{1}{\sqrt{2\Omega\hbar}}f=\frac{\partial \Im{\rm m}(z_\alpha \tilde{z}^*)}{\partial t},
\eeq
so that
\beq 
\int_0^{t_f} dt\, \Re{\rm e} (\tilde{z}) f=0, 
\label{zero}
\eeq
since $z_\alpha(t_b)=0$. 
With these results, we rewrite Eq. (\ref{phased}) as 
\beq
\phi_d=\frac{1}{\hbar}\int_0^{t_f} dt \left[\alpha+\sqrt{\frac{2\hbar}{\Omega}} \Re{\rm e}({\tilde{z}})\right]f=\frac{1}{\hbar}\int_0^{t_f} dt f \alpha. 
\eeq
Therefore, the geometric phase $\phi_g$ is minus the total phase,
\beq
\label{phase1}
\phi_g=\phi-\phi_d=-\frac{1}{2\hbar}\int_0^{t_f} dt f\alpha=-\phi.
\eeq
It is interesting to use the  phase-space trajectory in the rotating frame $z_r=e^{i\Omega t} z_g=X_r+iP_r$
to write ${f\la x\ra}/{\hbar}=\sqrt{\frac{2}{\hbar\Omega}}\Re{\rm e}(z_g)=2\Im{\rm m}(\frac{dz_r}{dt}z_r^*)=4d{\cal A}/dt$, where $d{\cal A}$ is the differential of 
area swept in the rotating phase space, $d{\cal A}/dt=\frac{X_r}{2}\frac{dP_r}{dt}-\frac{Pr}{2}\frac{dX_r}{dt}$. Thus    
Eq. (\ref{phased}) becomes
\beq
\label{phase2}
\phi_d=4{\cal A}. 
\eeq
Consequently, $\phi_g=-2{\cal A}$, and $\phi=2{\cal A}$. The area is equal for all trajectories (values of $z_g(0)$) due to Eq. (\ref{zero}),  
so it may  be calculated using $z_g(0)=0$, i.e., the simple particular solution $z_g=z_\alpha$. 
{Eqs. (\ref{phase1}) or (\ref{phase2}) are known results \cite{Garcia-Ripoll2005} but they are relevant for our work, so we have rederived 
them without using coherent states or a concatenation of displacement operators \cite{Leibfried2003}.  This is convenient when 
expressing the wave function  as a superposition in an orthonormal basis.}

\section{Invariant-based inverse Hamiltonian design\label{III}}
The results of the previous section may now be combined to inverse engineer the force. 
The Hamiltonian $H_{NM}$ involves the two modes so that superscripts or subscripts have to be added to 
the functions of the previous section to denote the mode. 

We assume that forces vanish at the boundary times $t_b=0,t_f$, $F_1(t_b)=F_2(t_b)=0$. 
Thus $f_\pm(t_b)=0$. 
In the rotating frame $\psi_I(t)=e^{iH_{NM}^0 t/\hbar}\psi_{NM}(t)$, where 
\beqa
H_{NM}^0 &=& H_+^0+H_-^0,
\nonumber\\
H_\pm^0&=&\frac{{\sf p}_\pm^2}{2} + \frac{1}{2}\Omega_\pm^2 {\sf x}_\pm^2,
\eeqa
so that 
\beq
\psi_I(t_f)=e^{-i[G_-(t_f)+G_+(t_f)]}\psi_I(0).
\eeq
Thus, the phase we are interested in for a given configuration is 
\beqa
\label{phase}
\phi(t_f)&=&-[G_+(t_f)+G_-(t_f)]
\nonumber\\
&=&-\frac{1}{2\hbar}\int_0^{t_f}\!\! dt' (\dot{\alpha}_+^2+\dot{\alpha}_-^2-\Omega_+^2\alpha_+^2-\Omega_-^2\alpha_-^2)
\nonumber\\
&=&\frac{1}{2\hbar}\int_0^{t_f}\!\! dt' (f_+\alpha_++f_-\alpha_-),
\eeqa
where the last step was done integrating by parts. An alternative double-integral expression used in Appendix \ref{AppendixB} is shown in Appendix \ref{AppendixC}.
The inverse strategy is to design the $\alpha_\pm$ consistently with the boundary conditions, leaving free parameters
that are fixed to produce the desired phase. The following section shows this in detail for equal masses.  
\section{Equal mass ions\label{IV}}
For  two equal-mass ions,   $m=m_1=m_2$, $\omega=\omega_1=\omega_2$, $a_+=-b_+=a_-=b_-=1/\sqrt{2}$, 
$\Omega_-=\omega$ (center of mass mode), and $\Omega_+=\sqrt{3}\omega$ (stretch mode). 
This implies, see Eq. (\ref{hanm}),  that 
\beq
\label{fgenequalm}
f_\pm=\frac{\pm F_2-F_1}{\sqrt{2m}},
\eeq
and $F_1$ and $F_2$ are defined as $F_i=\sigma_i^zF(t)$ (see the general case in Appendix \ref{AppendixB}), so that  
the following values are found
\beqa
\label{NMforceseq}
f_+(P)&=&f_-(A)=0,
\nonumber\\
f_-(\uparrow\uparrow)&=&f_+(\uparrow\downarrow)=-2F/\sqrt{2m},
\nonumber\\
f_-(\downarrow\downarrow)&=&f_+(\downarrow\uparrow)=2F/\sqrt{2m}, 
\eeqa
where $P$ stands for parallel spins, and $A$ for antiparallel ones. 
If both ions have the same spin, then $f_+(P)=0$ and no stretching is induced, 
but the center-of-mass $(-)$ mode is transiently excited. 
In that case, $\alpha_+(P)=0$ and 
%
\beq
\label{alm}
\alpha_-(\uparrow\uparrow)=-\alpha_-(\downarrow\downarrow),
\eeq
according to Eqs. \eqref{auxiliarya} and the established boundary conditions. 
For  opposite  spins $\alpha_-(A)=0$, and only the stretching (+) mode is transiently excited. 
In that case 
\beq
\label{alp}
\alpha_+(\uparrow\downarrow)=-\alpha_+(\downarrow\uparrow).
\eeq
%
The phase (\ref{phase}) takes two possible forms,
\beqa
\phi(P)&=& 
\frac{1}{\hbar}\int_0^{t_f} dt' \frac{-F}{\sqrt{2m}} \alpha_-(\uparrow\uparrow),
\nonumber\\
\phi(A)&=& 
\frac{1}{\hbar}\int_0^{t_f} dt' \frac{-F}{\sqrt{2m}} \alpha_+(\uparrow\downarrow).
\eeqa
%
%
To inverse engineer the phase 
we use the ansatz for $\alpha_+(\uparrow\downarrow;t)$ as a sum of Fourier cosines, with enough parameters to satisfy all boundary conditions,
\beq
\label{ansatzequal}
\alpha_+(\uparrow\downarrow; t)=a_0+\sum_{n=1}^4a_i\cos\left[\frac{(2n-1)\pi t}{t_f}\right].
\eeq
This is an odd function of $(t-t_f/2)$ which implies that $\ddot{\alpha}_+(\uparrow\downarrow; t)$, and thus 
$f_+(\uparrow\downarrow; t)$ are odd functions too
with respect to the middle time of the process $t_f/2$.     
The parameters $a_0$, $a_1$, and $a_2$ are fixed to satisfy the corresponding boundary conditions for $\alpha_+(\uparrow\downarrow)$ in Eqs. \eqref{conds1a} and \eqref{conds2a},  
\beqa
a_0&=&0,
\nonumber\\
a_1&=&2a_3+5a_4,
\nonumber\\
a_2&=&-3a_3-6a_4.
\eeqa
%
%
\begin{figure}[t]
\begin{center}
\includegraphics[width=8cm]{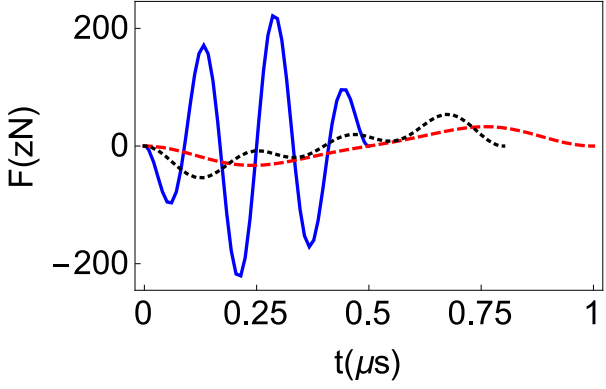}
\caption{\label{Forces}(Color online)
$F(t)$ for two $^9$Be$^+$ ions in a trap with frequency $\omega/2\pi=2$ MHz. 
$t_f=0.5$ $\mu$s (solid blue line),  $t_f=0.8$ $\mu$s (dotted black line), and  $t_f=1$ $\mu$s (dashed red line).  The forces on each ion are state-dependent, $F_i=\sigma_i^zF(t)$, $i=1,2$.}
\end{center}
\end{figure}
%
We get $f_+(\uparrow\downarrow; t)$ from Eq. \eqref{auxiliarya},  $f_+(\uparrow\downarrow; t)=\ddot{\alpha}_+(\uparrow\downarrow; t)
+\Omega_+^2\alpha_+(\uparrow\downarrow; t)$. Due to the boundary conditions, $f_+(0)=f_+(t_f)=0$. 
As $f_-(\uparrow\uparrow)=f_+(\uparrow\downarrow)$, we may solve Eq. \eqref{auxiliarya} for  $\alpha_-(\uparrow\uparrow; t)$ 
satisfying $\alpha_-(\uparrow\uparrow; t_b)=0$, which, as can be seen in Eq. \eqref{auxiliarya}, will be different from 
$\alpha_+(\uparrow\downarrow;t)$ because the frequencies of both normal modes are different ($\Omega_+ \neq \Omega_-$), although the 
forces are equal
[$\ddot{\alpha}_-(\uparrow\uparrow;t_b)=0$ is automatically 
satisfied since $f_-(\uparrow\uparrow, t_b)=0$]. The expression is rather lengthy but can be considerably simplified by imposing as well 
$\dot{\alpha}_-(\uparrow\uparrow; t_b)=0$. This fixes $a_3$ as  
%
%
\begin{figure}[t]
\begin{center}
\includegraphics[width=4cm]{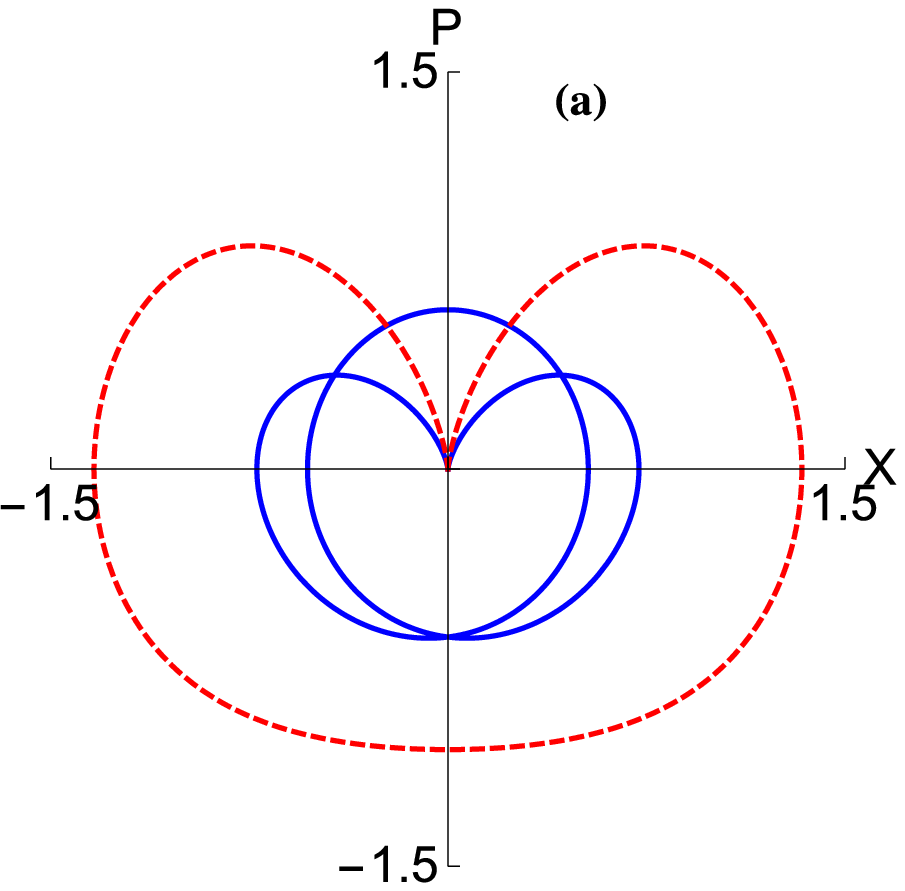}
\includegraphics[width=4cm]{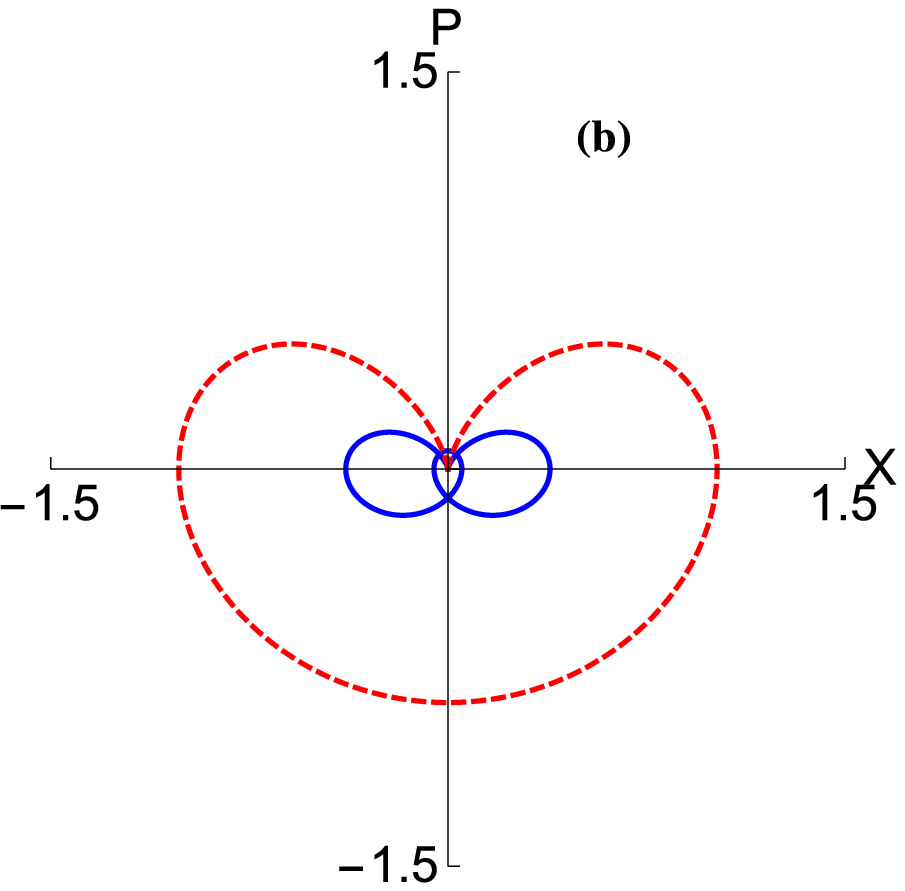}
\includegraphics[width=4cm]{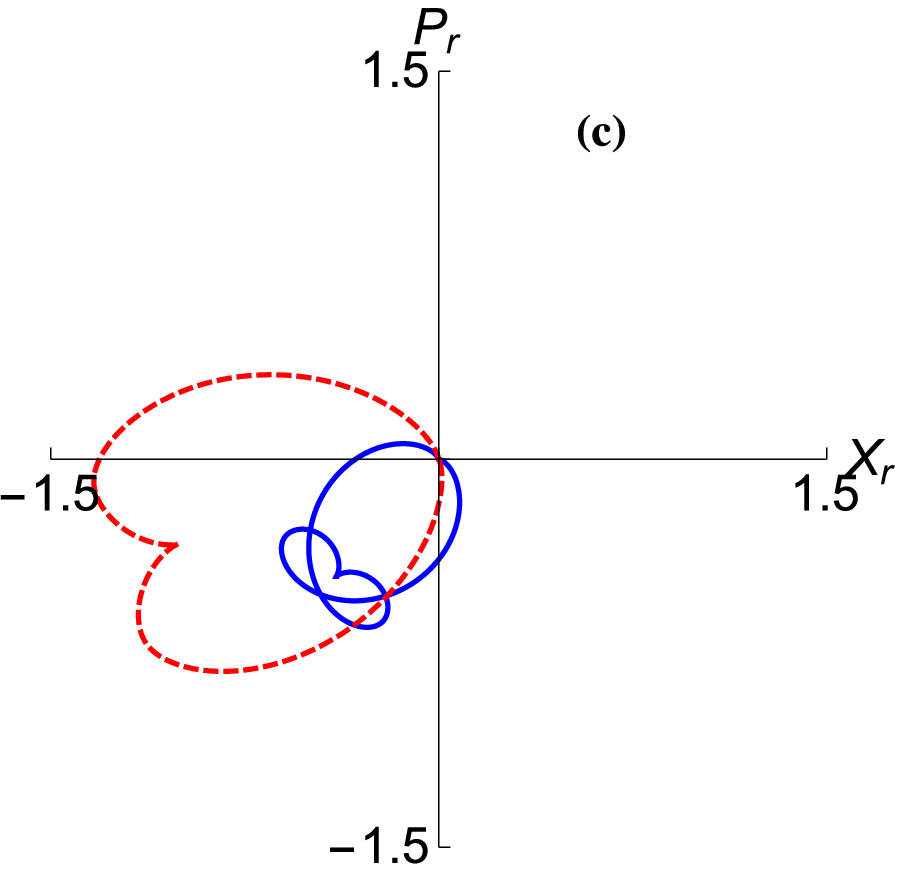}
\includegraphics[width=4cm]{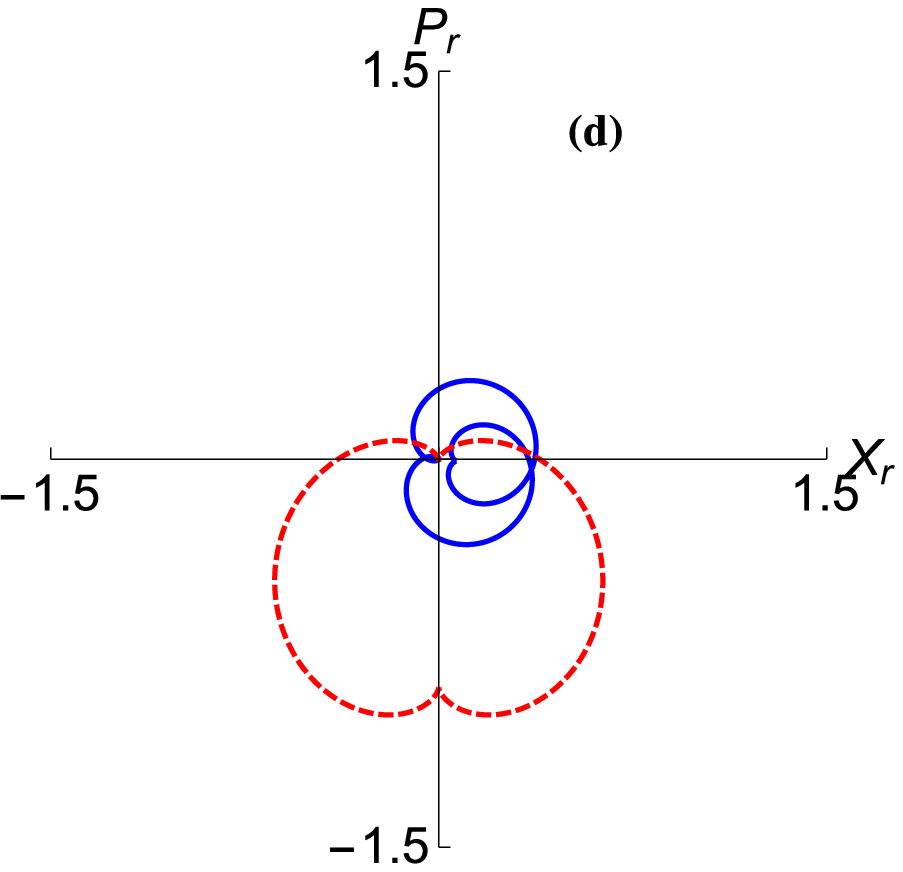}
\caption{\label{quadratures}(Color online) 
Parametric plots of the quadratures,  
$X=\sqrt{\frac{\Omega_\pm}{2\hbar}}\alpha_\pm$ and $P=\sqrt{\frac{1}{2\hbar\Omega_\pm}}\dot{\alpha}_\pm$.
The quadratures in the rotating frame are defined as $X_r=\Re{\rm e}(e^{i \Omega_\pm t} Z)$,   $P_r=\Im{\rm m}(e^{i \Omega_\pm t} Z)$, 
where $Z=X+iP$. 
The solid blue lines represent the stretch $(+)$ mode for antiparallel spins and the dashed red lines the 
center-of-mass $(-)$ mode for parallel spins. (a) $t_f=0.8$ $\mu$s, (b) $t_f=1$ $\mu$s, (c)  $t_f=0.8$ 
$\mu$s in the rotating frame, and (d) $t_f=1$ $\mu$s in the rotating frame. The other parameters are chosen as in 
Fig. \ref{Forces}.
}
\end{center}
\end{figure}
%
\beq
\label{a3equal}
a_3=-\frac{5a_4(25\pi^2-t_f^2\omega^2)}{49\pi^2-t_f^2\omega^2}.
\eeq
At this point  $\alpha_+(\uparrow\downarrow; t)$ and $\alpha_-(\uparrow\uparrow; t)$ are 
left as functions of the parameter $a_4$, 
\beqa
\label{alphas}
\alpha_+(\uparrow\downarrow; t) &=& \frac{11 \pi^2+t_f^2 \omega^2 +(49 \pi^2-t_f^2\omega^2)\cos{\frac{2 \pi t}{t_f}}}{49\pi^2-t_f^2\omega^2}
\nonumber\\
&\times& 32a_4\cos{\frac{\pi t}{t_f}}\sin^4{\frac{\pi t}{t_f}},
\nonumber\\
\alpha_-(\uparrow\uparrow; t) &=& \frac{11 \pi^2+3t_f^2 \omega^2 +(49 \pi^2-3t_f^2\omega^2)\cos{\frac{2 \pi t}{t_f}}}{49\pi^2-t_f^2\omega^2}
\nonumber\\
&\times& 32a_4\cos{\frac{\pi t}{t_f}}\sin^4{\frac{\pi t}{t_f}}.
\eeqa
These are both odd functions with respect to $(t-t_f/2)$ and guarantee a vanishing final excitation in the two modes. 
They also have vanishing third derivatives at the time boundaries and thus imply the continuity in the 
force derivative at time boundaries, i.e., $\dot{F}(t_b)=0$.  
Note that $-\frac{1}{\hbar}\int_0^{t_f} dt' 
[\tilde{f}(A)-\tilde{f}(P)]$ vanishes, see Eq. (\ref{hanm}), since $\tilde{f}(P)=0$ and $\tilde{f}(A)$ is also an odd function of $t-t_f/2$. 

The differential phase  
takes the form, see Eq. (\ref{phase}),  
\beqa
\Delta\phi&\equiv&2[\phi(A)-\phi(P)]
\nonumber\\
&=&\frac{2}{\hbar}\int_0^{t_f} dt' \frac{F}{\sqrt{2m}}[\alpha_-(\uparrow\uparrow)-\alpha_+(\uparrow\downarrow)].
\label{integral}
\eeqa
With the expressions \eqref{alphas} for  $\alpha_+(\ua\da)$ and $\alpha_-(\ua\ua)$ the integral can be solved to give  
\beq
\label{varphiequal}
\Delta\phi=\frac{12 a_4^2 t_f \omega^2(-2051 \pi^4+476 \pi^2 t_f^2 \omega^2-33 t_f^4 \omega^4)}
{\hbar(-49 \pi^2+t_f^2 \omega^2)^2}.
\eeq
Setting $\Delta\phi=\gamma$, the last free parameter is fixed as
\beqa
\label{a4}
a_4&=&\pm\frac{1}{\omega}(-147\pi^2+3t_f^2\omega^2)\sqrt{\frac{\hbar}{6t_f}}
\nonumber\\
&\times&
\left[\frac{\gamma/2}{-2051 \pi^4 +476 \pi^2 t_f^2 \omega^2-33 t_f^4 \omega^4}\right]^{1/2}.
\eeqa
%
%
%
\begin{figure}[t]
\begin{center}
\includegraphics[width=8cm]{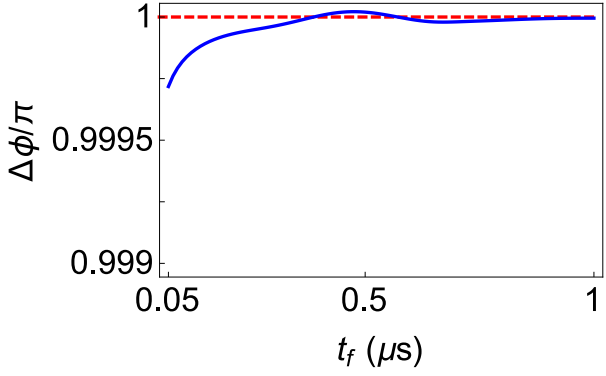}
\caption{\label{numericphase}(Color online)
$\Delta\phi$ 
for an exact evolution with the Hamiltonian in Eq. (\ref{Hamiltonian}),   and the 
target value of this phase (dashed red line). Same parameters as in Fig. \ref{Forces}.
The initial motional state is the ground state.}
\end{center}
\end{figure}
%
The polynomial denominator in the last term is negative for all $t_f$ 
(there are no real roots) so, to get a real $a_4$, $\gamma$ must be chosen as a negative number. 
We choose $\gamma=-\pi$ to implement the gate (\ref{gate2}). 
There are real solutions for 
$a_4$ for all $t_f$, no matter how small $t_f$ is. In this sense there is no fundamental lower bound for the method, 
as long as the small amplitude and Lamb-Dicke approximations are valid.  
As for the sign alternatives in $a_4$,  the different choices imply sign changes for the $\alpha$ and the forces.
We, hereafter and in all figures,  choose the positive sign. 
The resulting force takes the form
\beqa
\label{eqionsanalyticalforce}
F(t)&=&\frac{g_1(t_f)+g_2(t_f)\cos\left(\frac{2\pi t}{t_f}\right)+g_3(t_f)\cos\left(\frac{4\pi t}{t_f}\right)}{t_f^2\sqrt{2051\pi^4t_f\omega^2-476\pi^2t_f^3\omega^4+33t_f^5\omega^6}}
\nonumber\\
&\times&2\sqrt{2\pi\hbar m}\cos\left(\frac{\pi t}{t_f}\right)\sin^2\left(\frac{\pi t}{t_f}\right),
\label{Fexplicit}
\eeqa
where 
\beqa
g_1(t_f)&=&3(401\pi^4-36\pi^2t_f^2\omega^2+3t_f^4\omega^4), 
\nonumber\\
g_2(t_f)&=&-4(181\pi^4-76\pi^2t_f^2\omega^2+3t_f^4\omega^4), 
\nonumber\\
g_3(t_f)&=&2401\pi^4-196\pi^2t_f^2\omega^2+3t_f^4\omega^4,
\eeqa
%
which is shown in Fig. \ref{Forces} for different values of $t_f$. (All simulations in this section are for two 
$^{9}$Be$^+$ ions and a trap frequency $\omega/(2\pi)=2$ MHz.) 
The results are qualitatively similar to those found in \cite{Garcia-Ripoll2005}  (also the asymptotic behaviour for short operation times, $F\sim t_f^{-5/2}$) with a very different numerical method, but in our case the expression of $F$ is explicit, has a continuous envelope, and the derivatives vanish at the edges, adding stability.

With this force, the trajectories of $\alpha_+(A)$,  and $\alpha_-(P)$, see Eqs. (\ref{alm},\ref{alp}), 
are given in Fig. \ref{quadratures} for two given times $t_f$ in a dimensionless 
(quadrature) phase space, and in the rotating frame (the phase is twice the area swept in the rotating frame, see Sec. \ref{II} \cite{Garcia-Ripoll2005}). 
If the initial state is the ground state, they describe, respectively, the dynamics of the stretch mode for antiparallel spins
and the center of mass mode for parallel spins.  Notice that the trajectories 
lead to larger phase space amplitudes for  shorter times.

%
\begin{figure}[t]
\begin{center}
\includegraphics[width=8cm]{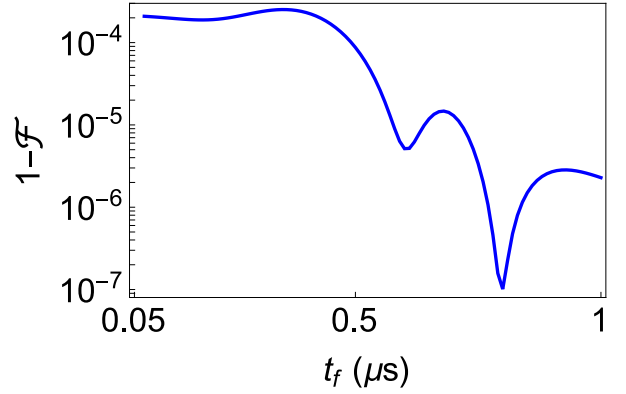}
\caption{\label{fidel}(Color online)
Worst case infidelity vs final time, which is realized for the initial state $|\uparrow\downarrow\ra$, see Eq. \eqref{worstfidelity}. 
Same parameters as in Fig. \ref{Forces}.}
\end{center}
\end{figure}
%

The phases within the harmonic (small amplitude) approximation are exact by construction for arbitrarily short times, 
but we should compare them with the phases when  
the system is driven by the full Hamiltonian (\ref{Hamiltonian}) that contains the anharmonic Coulomb interaction.   
To that end we solve numerically the Schr\"odinger equation with the Hamiltonian (\ref{Hamiltonian}) by using the ``Split-Operator Method" \cite{Kosloff1988}. 
First, we fix the initial state as the ground state of the system $|{\sf{\Psi}}_0\rangle$,
which is found by making an initial guess and evolving it in imaginary time \cite{Kosloff1986}. 
Then, the Split-Operator method is applied in real time to get the evolution of the wave function $|\sf{\Psi}_t\rangle$. 
Phases are much more sensitive than populations to numerical errors, so we need a much shorter time step than the one 
usually required until the results converge. 
At the final time, the overlap ${\cal S}=\langle{\sf{\Psi}}_0| {\sf{\Psi}}_{t_f}\rangle$ between the initial and the final state, 
which depends on the spin configuration, is calculated.  
The phase of the overlap is defined as $\varphi_f=\arg{{\cal S}}\in [0,2\pi)$. In the quadratic approximation this includes a global term $-(\Omega_++\Omega_-)t_f/2$, see Eq. (\ref{phasen}), absent in the rotating frame, that is canceled by  calculating the phase differential between antiparallel and parallel spins, $2[\varphi_f(\uparrow\downarrow)-\varphi_f(\uparrow\uparrow)]$, displayed in Fig. \ref{numericphase}.  
The corresponding infidelities, $1-|{\cal S}|^2\cos^2(\Delta\phi-\pi)=1-\mathcal{F}$, are shown in Fig. \ref{fidel} for the worst possible case, which is realized for an initial state with antiparallel spins (see Appendix \ref{AppendixD}). 
The numerical results agree with the ideal result of the quadratic 
approximation reasonably well at least up to operation times ten times smaller than an oscillation period $2\pi/\omega$,
i.e. $0.05$ $\mu$s for the parameters considered. 
The approximation may hold for even shorter times, but they are very demanding computationally.

\begin{figure}[t]
\begin{center}
\includegraphics[width=4cm]{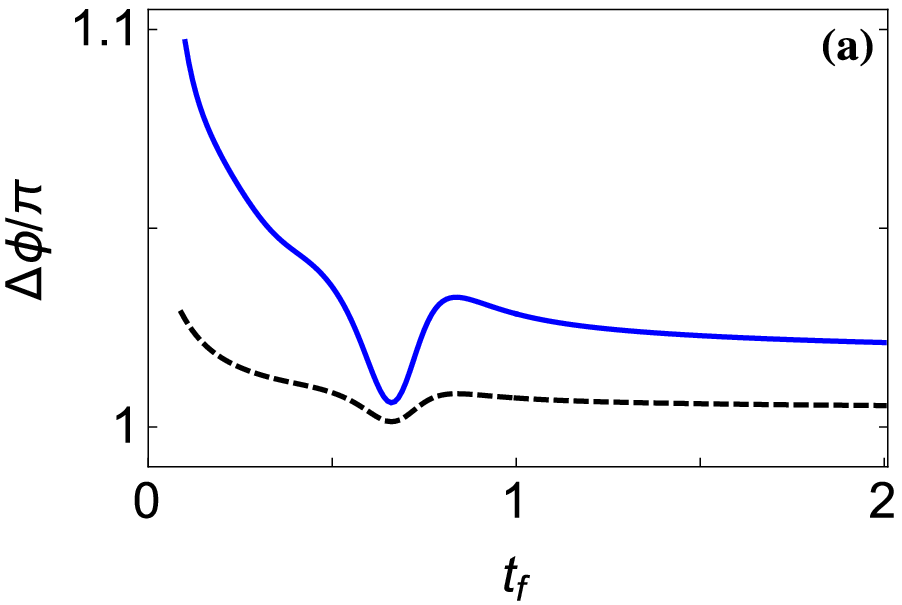}
\includegraphics[width=4cm]{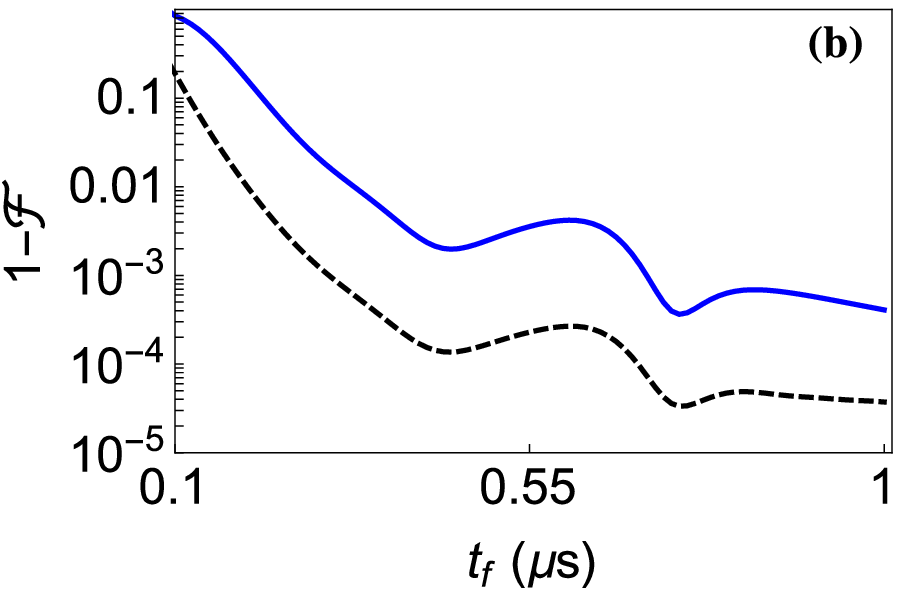}
\includegraphics[width=4cm]{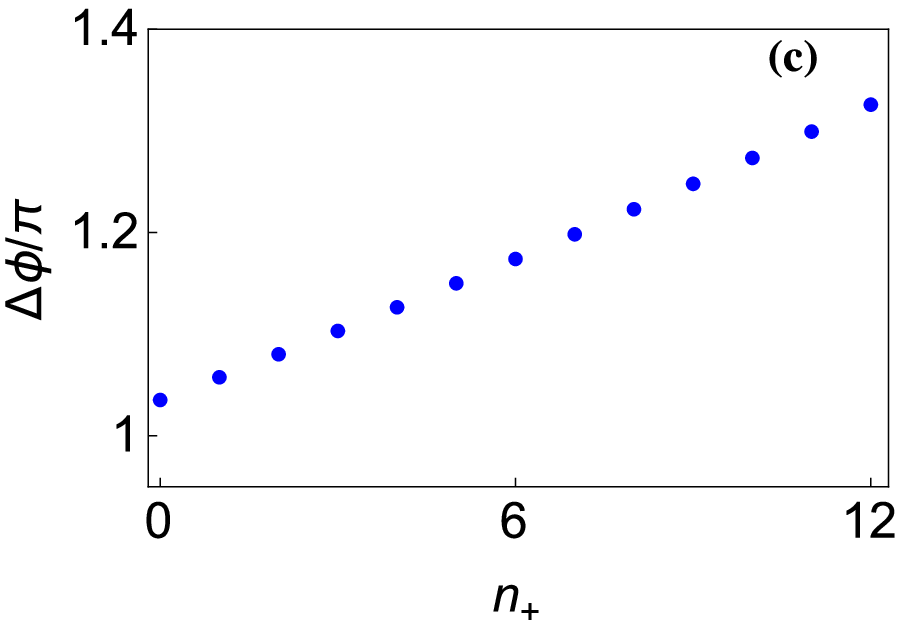}
\includegraphics[width=4cm]{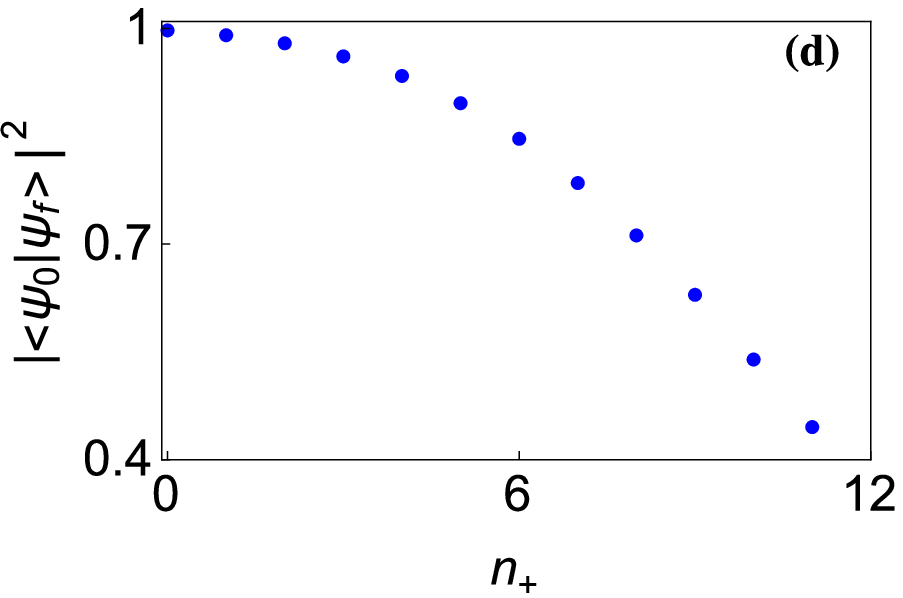}
\caption{\label{exc}(Color online)
Simulation of 2 $^9$Be$^+$ ions with trap frequency $\omega/(2\pi)=2$ MHz. Instead of homogeneous forces more realistic $x$-dependent forces
$F_i(t)\sin\left(\Delta k x+\pi/2\right)$ are applied. 
In (a) and (b) the initial motional state is the ground state.  $\Delta k=8.67\times 10^6$ m$^{-1}$: solid (blue) line (ions separated by 8 
lattice periods at equilibrium);  $\Delta k=4.33\times 10^6$ m$^{-1}$: dashed (black) line (ions separated by four lattice periods). In (a) we display the final phase vs the final time. In (b) the worst case infidelity (realized for antiparallel spins). In (c) and (d) the phase 
and worst case fidelity (corresponding to antiparallel spins) for different initial excited states are depicted, 
for a time $t_f=0.5$ $\mu$s and the ions separated by 
eight lattice periods.}
\end{center}
\end{figure}

A different type of stability check is displayed in Fig. \ref{exc}, where a realistic $x$-dependent sinusoidal force 
on each ion $F_i(t)\sin\left(\Delta k x+\pi/2\right)$ is considered instead of the homogeneous one. This force comes about because
of the finite wavelength of the lasers used to generate the  forces \cite{Leibfried2003}. Close to the ground state, 
the motional wave function of the ion only overlaps with a small part of the optical wave which can then be 
approximated as having a constant gradient over the wave function (Lamb-Dicke approximation). 
In more excited motional states, this approximation breaks down and the sinusoidal shape 
of the light wave has to be taken into account. For driving a phase gate, the wave-vector difference $\Delta k$ is adjusted so 
that the forces at the equilibrium positions $\pm x_0/2$ are the $F_i(t)$, with an integer number of periods $2\pi/\Delta k$ among them.  
{$\Delta k$ can be adjusted by changing the direction(s) of the beam(s) in laser-based experiments.}
We choose $\Delta k$ 
so that the ions in the equilibrium position for the frequency $\omega/(2\pi)=2$ MHz are placed in extrema of the sine function. 
In Fig. \ref{exc} (a) and (b) we depict the differential phase and worst case fidelity versus $t_f$ for this $x$-dependent force, starting in the motional ground state and performing the evolution for the full Hamiltonian as described in the previous paragraph. 
The two curves correspond to the ions being eight periods apart
at equilibrium, similar to \cite{Leibfried2003},  or four periods apart.    
As expected, the results degrade for very short times faster than for the ideal homogeneous case represented in Fig. \ref{fidel} 
since the ions explore a broader region where the forces deviate significantly from $F_i(t)$. 
The range of validity of the ideal results (the ones for a homogeneous force) in the limit $t_f\omega<<1$ 
is approximately given by $\frac{\Delta k}{\omega}\sqrt{\frac{\hbar}{t_f m}}<<1$, which 
may be found by estimating maximal amplitudes  of $\alpha_\pm$  in Eq. (\ref{alphas}),
using Eq. (\ref{x1x2}) to calculate deviations from equilibrium positions, and comparing them to half a lattice period 
$\pi/\Delta k$. Note that for eight periods the phase does not really converge to the ideal value even at 
longer times, when the deviation is quite small compared to the period of the force. The reason is that the wave function width also implies that the ions do not strictly experience a homogeneous force, which can lead to squeezing of the state of motion rather than just a coherent displacement.  Starting with the ground-state  wavefunction in the harmonic approximation (i.e., a product of ground-state wavefunctions for each mode), 
the width of the position of one ion is $\Delta x= \frac{1}{2} (1+1/\sqrt{3})^{1/2} \sqrt{\hbar/(m \omega)}$, see 
the Appendix \ref{AppendixE}, 
which should be compared to  $\pi/\Delta k$. For the parameters in Fig. \ref{exc} (a,b) the ratios $\Delta x\Delta k
/\pi$ are 0.04 (eight 
oscillations) and 0.02 (four oscillations).        

 In Figs. \ref{exc} (c) and (d), the evolved state begins in some excited Fock state
so that the Lamb-Dicke approximation breaks down more easily. 
We only consider excitations of the stretch mode, $|n_-=0,n_+\ra$,  as the full Hamiltonian
only has non-zero cubic terms for this mode.

\begin{figure}[t]
\begin{center}
\includegraphics[width=8cm]{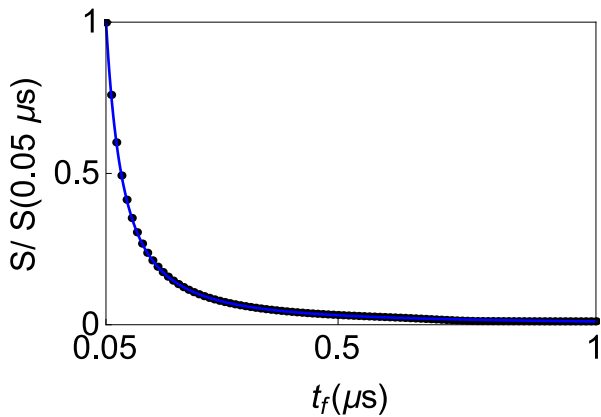}
\caption{\label{se}(Color online)
{$S=\int_0^{t_f} {dt |F(t)|}$ (dots).  $F(t)$ is designed for equal mass ions ($^9$Be$^+$ ions) according to Eq. (\ref{Fexplicit}) 
for a trap frequency $\omega/(2\pi)=2$ MHz. 
The solid line is a fit proportional to $t_f^{-3/2}$.} 
}
\end{center}
\end{figure}

We also study the scaling with $t_f$ of spontaneous emission due to the transitions induced by intense 
off-resonant fields. The transition rate will be proportional to the intensity 
of the field, and to the effective potential acting on the ions, i.e., to $|F|$. In Fig. \ref{se}
we have integrated this quantity over time for different values of $t_f$. Since $F\sim t_f^{-5/2}$
the result scales as $t_f^{-3/2}$. We have normalized the integral by the force $|F(0.05 {\mu}s)|$ as the scattering probability will depend on 
different factors which are not explicitly considered here such as $\Delta k$ or the detuning.
%
%
%

%
%
\begin{table}[t]
\centering
\setlength{\arrayrulewidth}{.1em}
\bigskip
\begin{tabular}{l  c  c  c }
\hline \\
&  $m$ (a.u.) & \hspace{0.3cm} $\omega/(2\pi)$ (MHz) & \hspace{0.3cm} $F_{max}$ (zN) \\ \\ \hline
\\
Be \hspace{0.7cm} & 9 & 2 & 223.89 \\ 
Mg & 24 & 2$\sqrt{\frac{9}{24}}$ & 365.61 \\ 
Ca & 40 & 2$\sqrt{\frac{9}{40}}$ & 472.00 \\ 
Sr & 88 & 2$\sqrt{\frac{9}{88}}$ & 700.09 \\ 
Ba & 138 & 2$\sqrt{\frac{9}{138}}$ & 876.70 \\ \\ \hline
\end{tabular}
\bigskip
\caption{Maximum value of the force $F$ in Eq. \eqref{eqionsanalyticalforce} for different alkaline earth ions, and their corresponding mass and trap frequency. The parameters where chosen so that they all have the same spring constant. The calculated maximum force is for an operation time $t_f=0.5$ $\mu$s.}
\label{table3_1} 
\end{table}
Finally, in Table \ref{table3_1} we calculate the maximum value of the driving force in Eq. \eqref{eqionsanalyticalforce} for  different alkaline earth ion species.  Their potential performance for phase gates was analyzed in  
\cite{Staanum2004}.  

%
%
%
%
\section{Different masses\label{V}}
For different mass ions, $m_1=m$, $m_2=\mu~m$, and $u_0=m\omega_1^2=\mu m\omega_2^2$. In this case, due to their different structure, both ions will react to different laser fields, thus, $F_1$ and $F_2$ can in principle be designed independently, such that $F_1=\sigma_1^zF_a(t)$, $F_2=\sigma_2^zF_b(t)$
(more general cases are studied in Appendix \ref{AppendixB}), yielding 
\beqa
\label{fnoneq}
f_\pm(\uparrow\uparrow) &=& - f_\pm(\downarrow\downarrow)= \mp\frac{b_\mp}{\sqrt{m}} F_a \pm \frac{a_\mp}{\sqrt{\mu m}}F_b,
\nonumber\\
f_\pm(\uparrow\downarrow) &=&-f_\pm(\downarrow\uparrow)= \mp\frac{b_\mp}{\sqrt{m}} F_a \mp \frac{a_\mp}{\sqrt{\mu m}} F_b,
\eeqa
which, as in the previous section, implies that
\beqa
\alpha_\pm(\downarrow\uparrow) &=& -\alpha_\pm(\uparrow\downarrow),
\nonumber\\
\alpha_\pm(\downarrow\downarrow) &=& -\alpha_\pm(\uparrow\uparrow),
\eeqa
see Eqs. \eqref{hanm} and \eqref{auxiliarya},
so if the protocol is designed to satisfy the boundary conditions for the $\uparrow\downarrow$ and $\uparrow\uparrow$ configurations, it will automatically satisfy the conditions for the remaining configurations. 
Inversely, from Eq. (\ref{fnoneq}) and Eq. (\ref{abrela}), 
\beqa
F_a &=&-\sqrt{m} [a_-f_-(\uparrow\downarrow) + a_+f_+(\uparrow\downarrow)],
\nonumber\\
F_b &=& \sqrt{\mu m}[b_-f_-(\uparrow\downarrow) + b_+f_+(\uparrow\downarrow)].
\label{FaFb0}
\eeqa
The procedure to design the forces is summarized in the following scheme,
\beq
\label{solscheme}
\alpha_\pm(\uparrow\downarrow)\dashrightarrow f_\pm(\uparrow\downarrow) \dashrightarrow F_a, F_b \dashrightarrow f_\pm (\uparrow\uparrow) \dashrightarrow \alpha_\pm(\uparrow\uparrow).
\eeq
To start with, ansatzes are proposed for $\alpha_+(\uparrow\downarrow)$ and $\alpha_-(\uparrow\downarrow)$, 
%
\beqa
\label{ansatzdiff}
\alpha_+(\uparrow\downarrow) &=& a_0+\sum_{n=1}^4 a_n \cos\left[\frac{(2n-1)\pi t}{t_f}\right],
\nonumber\\
\alpha_-(\uparrow\downarrow) &=& 0.
\eeqa
It is also possible to design them so as to cancel $\alpha_+(\uparrow\downarrow)=0$ 
instead of $\alpha_-(\uparrow\downarrow)$, as discussed in Appendix \ref{AppendixF}. 
Similarly to the previous section, $a_0$, $a_1$, and $a_2$ 
are fixed to satisfy the  boundary conditions for $\alpha_+(\uparrow\downarrow)$ in Eqs. \eqref{conds1a} and \eqref{conds2a},  
\beqa
a_0=0, \quad a_1 = 2a_3+5a_4, \quad a_2 = -3a_3-6a_4.
\label{abs}
\eeqa
Introducing these ansatzes in Eq. \eqref{auxiliarya}, expressions for $f_\pm(\uparrow\downarrow;t)$ are found, 
in particular $f_-(\ua\da)=0$, and from these,  
expressions for the control functions $F_a(t), F_b(t)$ follow according to Eq. (\ref{FaFb0}).   
Since $\alpha_+(t)(\uparrow\downarrow)$ is an odd function of $(t-t_f/2)$,
the same symmetry applies to $F_a(t)$, $F_b(t)$, and to the spin-dependent forces $F_1$, $F_2$.  
Thus the time integral of $\tilde{f}$, see Eq. (\ref{hanm}), is zero  for different masses as well and does not contribute to the 
phase.    

%
%
\begin{figure}[t]
\begin{center}
\includegraphics[width=8cm]{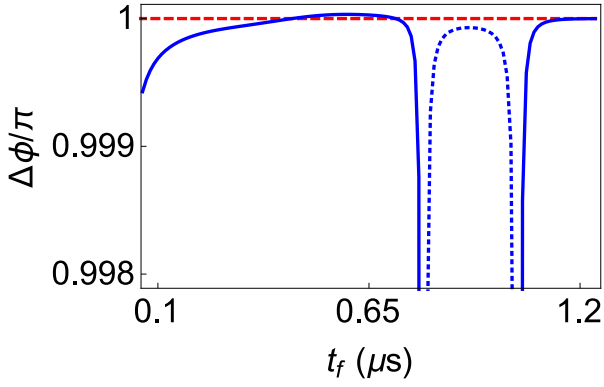}
\caption{\label{phasedif}(Color online)
Total final phase $\Delta\phi$ in Eq. \eqref{varphidifferent} vs the final time (solid and dotted blue lines) 
for an exact wave function evolving with the Hamiltonian in Eq. (\ref{Hamiltonian}), and the 
target value of this phase (dashed red line). 
The simulation is done for a $^9$Be$^+$ and a $^{25}$Mg$^+$ ion, initially in the motional ground state, 
within a trap of frequency $\omega_1/(2\pi)=2$ MHz.
At final times $t_f\sim 0.8$ $\mu$s and $1.03$ $\mu$s we change solutions, see the discussion below Eq. (\ref{70}).
The solid line is for $\gamma=-\pi$ and the dashed line for $\gamma=\pi$.}
\end{center}
\end{figure}
%
%
%
\begin{figure}[t]
\begin{center}
\includegraphics[width=4cm]{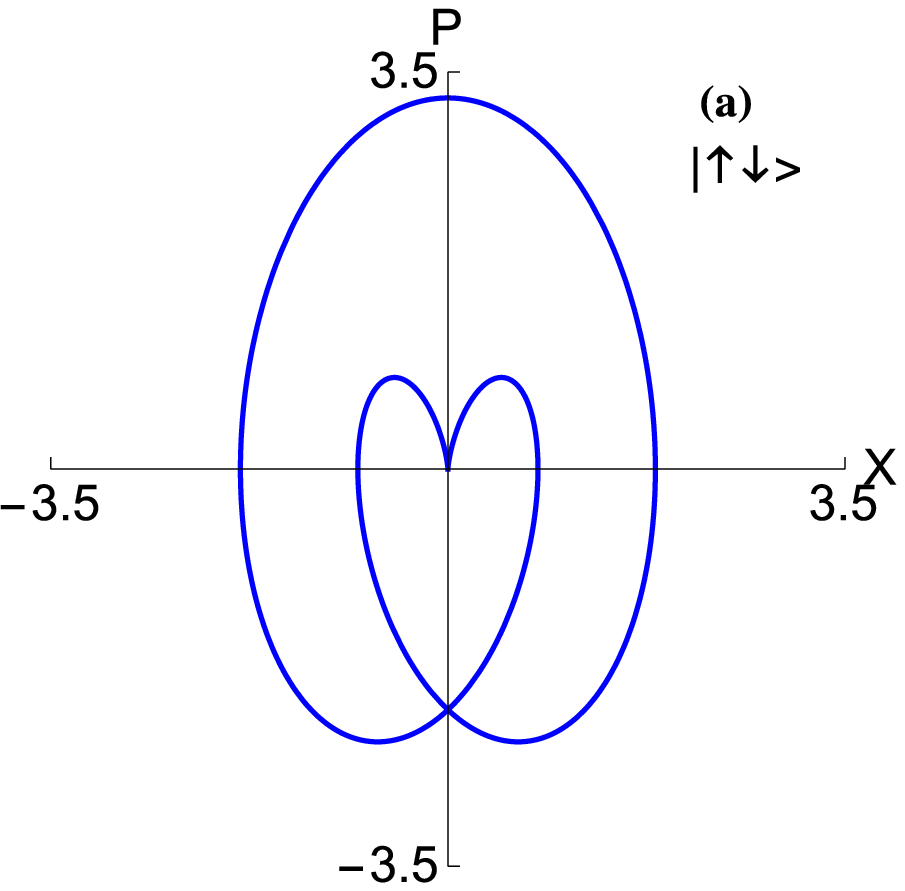}
\includegraphics[width=4cm]{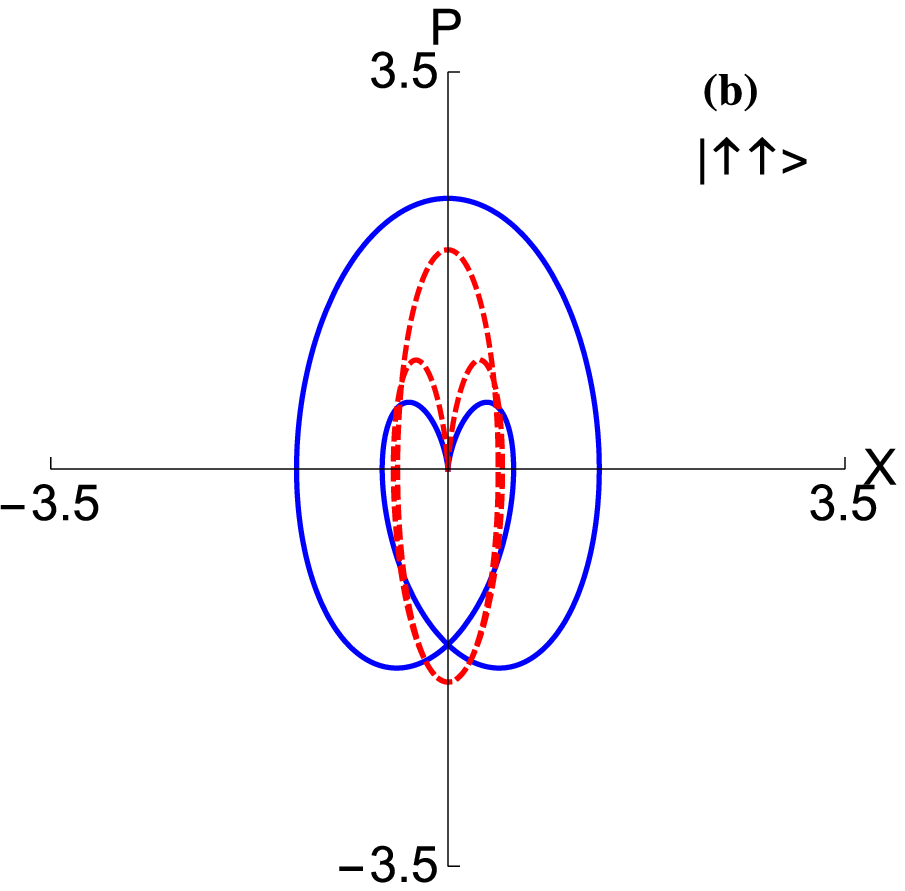}
\includegraphics[width=4cm]{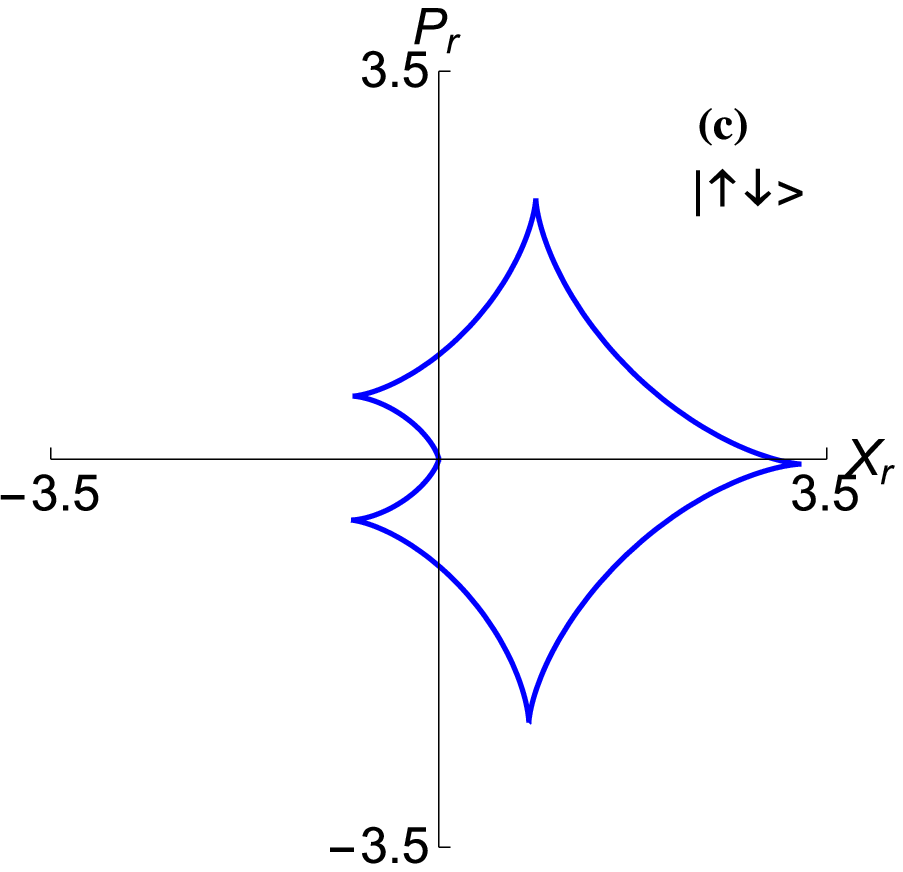}
\includegraphics[width=4cm]{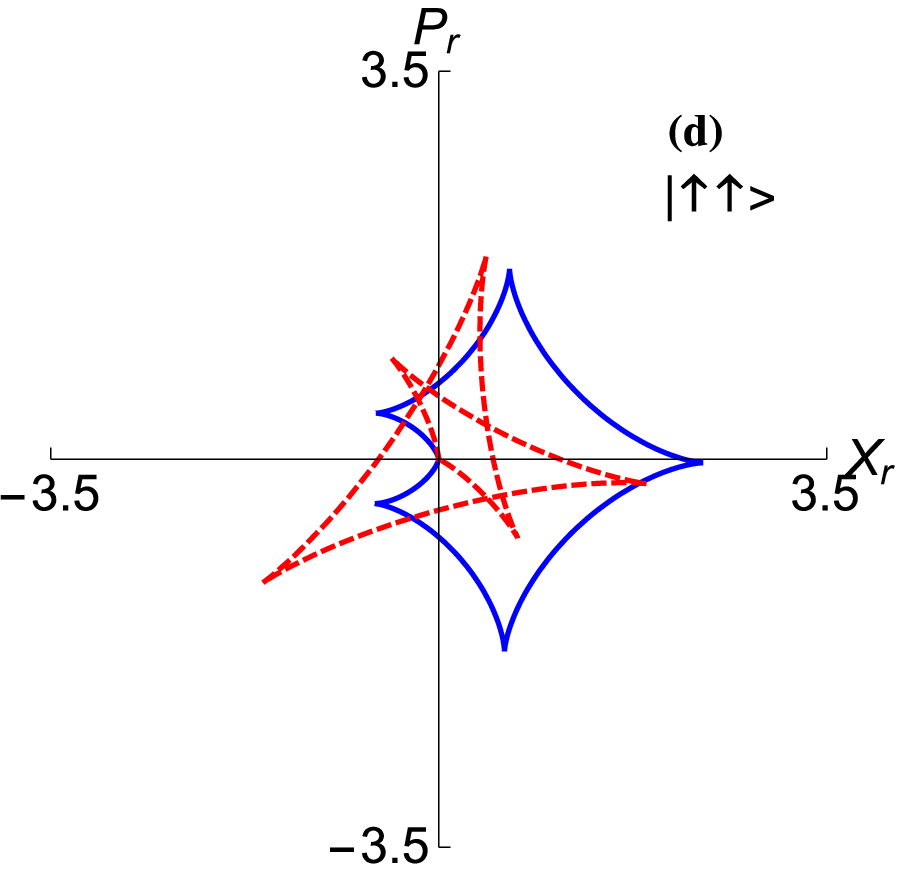}
\caption{\label{quadraturesdiff}(Color online) 
Parametric plots of the quadratures,  
$X=\sqrt{\frac{\Omega_\pm}{2\hbar}}\alpha_\pm$ and $P=\sqrt{\frac{1}{2\hbar\Omega_\pm}}\dot{\alpha}_\pm$.
The quadratures in the rotating frame are defined as $X_r=\Re{\rm e}(e^{i \Omega_\pm t} Z)$,   $P_r=\Im{\rm m}(e^{i \Omega_\pm t} Z)$, 
where $Z=X+iP$ at $t_f=0.5$ $\mu$s. 
The solid blue lines represent the stretch $(+)$ mode and the dashed red lines the 
center-of-mass $(-)$ mode. (a) and (c) represent the phase space trajectory for $|\uparrow\downarrow\ra$
and $|\downarrow\uparrow\ra$, in the normal and the rotating frame respectively, while (b) and (d) represent the 
phase space trajectories for $|\uparrow\uparrow\ra$, in the normal and rotating frames respectively. The other parameters are chosen as in 
Fig. \ref{phasedif}.
}
\end{center}
\end{figure}
%
%
Using the last line of Eq. \eqref{hanm},  the effective forces $f_\pm(\uparrow\uparrow)$ are found. Plugging these functions into Eq. \eqref{auxiliarya} we solve the differential equations imposing the boundary conditions $\alpha_\pm(\uparrow\uparrow ; t_b)=0$ to fix the integration constants. At this point the boundary conditions for 
$\ddot{\alpha}_\pm(\uparrow\uparrow; t_b)$ are automatically satisfied, and $\dot{\alpha}_\pm(\uparrow\uparrow ; 0)=\dot{\alpha}_\pm(\uparrow\uparrow ; t_f)$ by symmetry. Thus,  imposing 
that the first derivatives vanish 
at the boundary times,  $a_3$ is fixed as 
\beqa
\label{a3diff}
a_3 =\frac{-25 \pi^2 + t_f^2 \Omega_-^2}{49 \pi^2 - t_f^2 \Omega_-^2} 5 a_4.
\eeqa
%
%
Once the $\alpha_\pm$ are given for both configurations, such that they do not produce any excitation in the modes at the final time, 
the final phase difference is, as in the previous section,
\beqa
\label{varphidifferent}
\Delta\phi (t_f) &= &2[\phi (A)-\phi(P)]
\nonumber\\
&=& -\frac{1}{\hbar}\sum_{\mu=\pm}\int_0^{t_f}dt[\dot{\alpha}_\mu^2(\uparrow\downarrow)-\Omega_\mu^2\alpha_\mu^2(\uparrow\downarrow)]
\nonumber\\
&+& \frac{1}{\hbar}\sum_{\mu=\pm}\int_0^{t_f}dt [\dot{\alpha}_\pm^2(\uparrow\uparrow)-\Omega_\mu^2\alpha_\mu^2(\uparrow\uparrow)].
\eeqa
The integrals can be evaluated and give a function of $a_4$. This parameter is finally set by imposing some value to the phase difference, 
$\Delta\phi (t_f)=\gamma$,
\beqa
\label{a4diff}
a_4&=&\sqrt{\frac{\gamma \hbar(1+(-1+\mu)\mu) (-49\pi^2+t_f^2 \Omega_-^2)^2}{\Delta}},
\nonumber\\
\Delta&=&   6 \mu (\Omega_--\Omega_+)(\Omega_-+\Omega_+)\nonumber\\
&\times& t_f[2051\pi^4+11t_f^4\Omega_-^2\Omega_+^2-119\pi^2 t_f^2 (\Omega_-^2 + \Omega_+^2)].
\nonumber\\
\eeqa
The function $\Delta$ has zeros at
\beqa
\label{criticaltimes}
t_f^{(0)} &=& 0,\nonumber\\
t_f^{(1)} &=& \pm\pi \sqrt{\frac{119 (\Omega_-^2 +\Omega_+^2) -\delta}{22\Omega_-^2 \Omega_+^2}},\nonumber\\
t_f^{(2)} &=& \pm\pi \sqrt{\frac{119 (\Omega_-^2 +\Omega_+^2) +\delta}{22\Omega_-^2 \Omega_+^2}},
\eeqa
where 
\beq\label{70}
\delta= \sqrt{7(2023 \Omega_-^4 -8846 \Omega_-^2 \Omega_+^2 +  2023 \Omega_+^4)}.
\eeq
Considering only the positive times, 
in the intervals $\left(t_f^{(0)},t_f^{(1)}\right)$ and $t_f>t_f^{(2)}$, $\Delta$ is negative, 
so we chose  $\gamma=-\pi$ to make  $a_4$, and thus $F_a, F_b$,  real. 
In the interval $\left(t_f^{(1)},t_f^{(2)}\right)$ $\Delta$ is positive, so we can  choose $\gamma=\pi$.
%
%
\begin{figure}[t]
\begin{center}
\includegraphics[width=8cm]{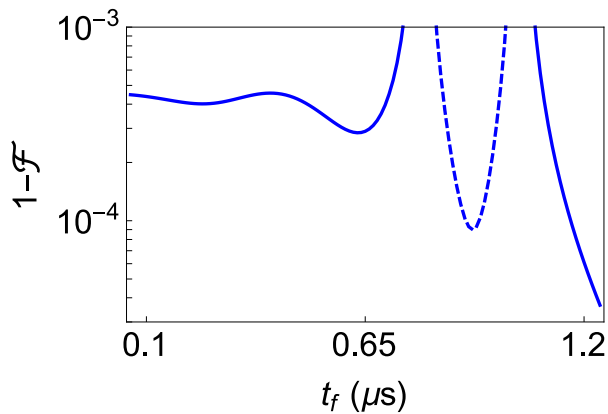}
\caption{\label{fiddif}(Color online)
Infidelity vs final time for the worst case, which corresponds to antiparallel spins, see Eq. \eqref{worstfidelity}. Same parameters as in Fig. \ref{phasedif}.
The solid line is for $\gamma=-\pi$ and the dashed line for $\gamma=\pi$.}
\end{center}
\end{figure}
%
%

The explicit expressions for the control functions are finally, from Eq. (\ref{FaFb0}),  
\beqa
\label{controlparamdiff}
F_a &=& \left[g_1^a+g_2^a\cos\left(\frac{2\pi t}{t_f}\right)+g_3^a\cos\left(\frac{4 \pi t}{t_f}\right)\right]
\nonumber\\ 
 &\times &
\frac{8 a_4 a_+ \sqrt{m}\cos\left(\frac{\pi t}{t_f}\right) \sin^2\left(\frac{\pi t}{t_f} \right)}{-49 \pi^2 t_f^2 + t_f^4 \Omega_-^2},
\nonumber\\
F_b &=& -\frac{b_+\sqrt{\mu}}{a_+}F_a,
\eeqa
where
\beqa
g_1^a &=& 3 [401 \pi^4 + t_f^4 \Omega_-^2 \Omega_+^2 - 9 \pi^2 t_f^2 (\Omega_-^2 + \Omega_+^2)], 
\nonumber\\
g_2^a &=& 4[-181 \pi^4 -  t_f^4 \Omega_-^2 \Omega_+^2 + 19 \pi^2 t_f^2 (\Omega_-^2 + \Omega_+^2)],
\nonumber\\
g_3^a &=& (49 \pi^2 - t_f^2 \Omega_-^2) (49 \pi^2 - t_f^2 \Omega_+^2).
\eeqa
%
$F_a,F_b$ diverge for the final times in Eq. \eqref{criticaltimes}, so these times must be avoided. 
{The positions of the divergences depend on the chosen ansatz.
In particular, for a polynomial, rather than cosine ansatz, the only divergence is at $t_f=0$. We have however kept the cosine ansatz 
as it needs fewer terms and it simplifies the results and the treatment of boundary conditions.}    

Figure \ref{phasedif} shows the phase  found numerically with the exact Hamiltonian 
for  $^9$Be (ion 1) and $^{25}$Mg (ion 2) in the Lamb-Dicke limit beginning in the ground motional state. 
Figure \ref{quadraturesdiff} shows the quadratures for such a protocol at final time $t_f=0.5$ $\mu$s, and Fig. \ref{fiddif} the worst case infidelities at final time, which, as in the previous section, correspond to initial states with antiparallel
spin (see Appendix \ref{AppendixD}). 
Around an oscillation period $2\pi/\omega_1=0.5$ $\mu$s, the results are slightly worse than in the previous section for equal mass ions, but still 
with a high fidelity.
For final times close to $t_f^{(1)}\sim 0.8$ $\mu$s and $t_f^{(2)}\sim 1.03$ $\mu$s
the solutions change, with a drop in the stability of the phase (Fig. \ref{phasedif}) and in the fidelity (Fig. \ref{fiddif}). The phase and fidelity improve and stabilize again for times $t_f> 1.03$ $\mu$s. 

In the limit were both ions are equal, $F_a=F_b=F$, and the results of the previous section are found consistently. 
\section{Discussion\label{VI}}
In this paper, we have designed simple and explicit protocols to perform fast and high fidelity phase gates with two trapped ions by using the 
invariant-based method to bypass adiabaticity. 
The scheme of the gate expands on methods that have been already tested in the laboratory. 
Experimentally, the state-dependent forces may be created by a standing wave  with time-varying intensity 
produced by two crossed laser beams whose amplitude is modulated following a smoothly designed trajectory to excite motion in both normal modes.
In the limit of small oscillations, we can use both a normal mode harmonic approximation and the Lamb-Dicke limit and apply the inverse-design  method assuming homogeneous forces. We have also numerically simulated the system dynamics and gate behavior without these approximations, namely, including the anharmonicity and the position dependence of the forces. Good fidelities are obtained at times around 1 $\mu$s, which is a significantly shorter time compared to the best experimental results so far and close to the center of mass oscillation period which was assumed to be 0.5 $\mu$s in this work. Moreover, state-of-the-art technology allows for higher trap frequencies than those used in our simulations, which should further improve the results.
Expressions for the forces have been found for different scenarios, specifically for equal or different masses, as well as for different proportionality factors between the spin-dependent forces.  

At present, technical limitations for the laser intensity and Raman-detuning will constrain
 the shortest gate times that can be reached. However, technical limitations can change, especially if there is motivation to push them. The goal of our manuscript is to also explore the possibilities that exist beyond what is presently doable.

Extensions of this work are possible in several directions. For example, the deviations from the ideal conditions 
may be taken into account to design the forces. The force design given here, based on symmetrical
trajectories   of the modes, gives already stable results 
(a vanishing first order correction to the phase) with respect to a systematic, 
homogeneous deviation of the forces, see Eq. (\ref{Delphasea}).  
Moreover the freedom offered by the approach 
may also be used to choose stable protocols with respect to different noises and perturbations, see e.g. 
\cite{Andreas,Lu,Lu2,Daems}.   
\section*{Acknowledgments}
We thank A. Ruschhaupt and J. Alonso for useful discussions.  
This work was partially supported by the Basque Government (Grant IT472-10), MINECO (Grant FIS2015-67161-P), and the program UFI 11/55 of UPV/EHU. 
D.L. and D.W.  acknowledge support by the Office of the Director of National Intelligence (ODNI) Intelligence Advanced Research Projects Activity (IARPA), ONR and the NIST Quantum Information Program.
M.P. and S.M.-G. acknowledge fellowships by UPV/EHU.

\appendix
\section{Generalization for an arbitrary force ratio\label{AppendixB}}
\subsection{Equal-mass ions\label{emi}}
In the main text we have studied state-dependent forces which are equal and opposite to each other for up and down spins, 
$F_i=\sigma_i^zF(t)$. However, depending on laser beam
polarization and specific atomic structure, different proportionalities among the two forces will arise. 
Let us consider a general force ratio
$F_i(\uparrow)=-c \tilde{F}(t)$ and $F_i(\downarrow)=-\tilde{F}(t)$, where $c$ is a constant. 
Then, for equal-mass ions, instead of Eq. (\ref{NMforceseq}) (corresponding to $c=-1$), we find, see Eq. (\ref{fgenequalm}),   
\beqa
\label{appendixforces}
f_+(\uparrow\uparrow) &=& f_+(\downarrow\downarrow)=0,
\nonumber\\
f_-(\uparrow\uparrow) &=& c f_-(\downarrow\downarrow)=\frac{2\tilde{F}c}{\sqrt{2 m}},
\nonumber\\
f_+(\uparrow\downarrow) &=& -f_+(\downarrow\uparrow)=-\frac{1-c}{\sqrt{2m}}\tilde{F},
\nonumber\\
f_-(\uparrow\downarrow) &=& f_-(\downarrow\uparrow)=\frac{1+c}{\sqrt{2m}}\tilde{F}.
\eeqa
%
To inverse engineer the forces we start choosing  the same ansatz for $\alpha_+(\downarrow\uparrow)$ as in Eq. \eqref{ansatzequal}.
$a_0$ through $a_2$ are also fixed in the same manner to satisfy the boundary conditions.
Using Eq. (\ref{auxiliarya}) this gives $f_+(\downarrow\uparrow;t)$ as a function of $a_3,\, a_4$, and in fact all other $f_\pm$ by scaling them according to the Eq. (\ref{appendixforces}).       
As in the main text, the same $a_3$ in Eq. \eqref{a3equal} guarantees that $\dot{\alpha}_\pm(t_b)=0$ for all spin configurations. Now, using Eq. \eqref{phase} we can 
write down the phase produced by each spin configuration. Individually, they depend on $c$ but, adding them all in  $\Delta\phi=\phi(\uparrow\downarrow)+\phi(\downarrow\uparrow)-\phi(\uparrow\uparrow)-\phi(\downarrow\downarrow)$, the dependence on $c$ is cancelled, as can be seen from Eq. \eqref{phaseforces} or Eq. (\ref{auxiliarya}) and Eq. (\ref{appendixforces}). Following the method described in the main text, imposing $\Delta\phi=\gamma$ fixes the remaining parameter $a_4$, so that the same expression in Eq. \eqref{a4} is found. 
Using Eqs. \eqref{NMforceseq} and \eqref{appendixforces}, the generic control function $\tilde{F}$ is simply proportional to that for $c=-1$ (see Eq. (\ref{Fexplicit})), 
\beq
\tilde{F}=\frac{2}{1-c}F.
\eeq
\subsection{Different masses}
Similarly, for different-mass ions in the generic case both ions could have different proportionality factors for the spin-dependent forces:  
\beqa
F_1(\uparrow)&=&-c_1\tilde{F}_a,\;\; F_1(\downarrow)=-\tilde{F}_a,
\nonumber\\
F_2(\uparrow)&=&-c_2\tilde{F}_b,\;\; F_2(\downarrow)=-\tilde{F}_b. 
\eeqa
Instead of Eq. (\ref{fnoneq}), 
the normal-mode
forces are now, see Eq. (\ref{hanm}), 
\beqa
f_\pm(\uparrow\uparrow) &=&\pm \frac{b_\mp}{\sqrt{m}} c_1\tilde{F}_a \mp \frac{a_\mp}{\sqrt{\mu m}} c_2\tilde{F}_b,\nonumber\\
f_\pm(\uparrow\downarrow) &=&\pm \frac{b_\mp}{\sqrt{m}} c_1\tilde{F}_a \mp \frac{a_\mp}{\sqrt{\mu m}} \tilde{F}_b,\nonumber\\
f_\pm(\downarrow\uparrow) &=& \pm \frac{b_\mp}{\sqrt{m}} \tilde{F}_a \mp \frac{a_\mp}{\sqrt{\mu m}} c_2\tilde{F}_b,\nonumber\\
f_\pm(\downarrow\downarrow) &=&\pm\frac{b_\mp}{\sqrt{m}} \tilde{F}_a \mp \frac{a_\mp}{\sqrt{\mu m}} \tilde{F}_b.
\label{allf}
\eeqa
%
This implies that the $\alpha_\pm$ are in general all different and the inverse scheme in Eq. \eqref{solscheme} is replaced by 
\beq
\alpha_\pm(\uparrow\downarrow)\dashrightarrow f_\pm(\uparrow\downarrow) \dashrightarrow 
\underbrace{
\tilde{F}_a, \tilde{F}_b \dashrightarrow  
\left\{
\begin{array}{cll}
f_\pm(\uparrow\uparrow) & \dashrightarrow & \alpha_\pm(\uparrow\uparrow)
\\
f_\pm(\downarrow\uparrow) & \dashrightarrow & \alpha_\pm(\downarrow\uparrow)
\\
f_\pm(\downarrow\downarrow) & \dashrightarrow & \alpha_\pm(\downarrow\downarrow).
\end{array}\right.
}_{\mbox{functions of} \,\, c_1,c_2}
\eeq
Using Eq. (\ref{abrela}) and Eq. (\ref{allf}) we may rewrite the control functions $\tilde{F}_a$ and $\tilde{F}_b$  as
\beqa
\label{FaFb}
\tilde{F}_a &=& \sqrt{m}[a_-f_-(\uparrow\downarrow)+a_+f_+(\uparrow\downarrow)]/c_1,
\nonumber\\
\tilde{F}_b &=& \sqrt{\mu m}[b_-f_-(\uparrow\downarrow)+b_+f_+(\uparrow\downarrow)].
\eeqa
As in the special case $c_1=c_2=-1$ of the main text, we use the ansatzes in Eq. \eqref{ansatzdiff} for $\alpha_\pm(\uparrow\downarrow)$,  
and the parameters in Eq. (\ref{abs}).  In particular $\alpha_-(\ua\da)=0$ and  
$f_-(\uparrow\downarrow)=0$, so $\tilde{F}_a$ and $\tilde{F}_b$ are proportional to each other, see Eq. (\ref{FaFb}), 
and thus all the $f_\pm$ are proportional to $f_+(\uparrow\downarrow)$ according to Eq. (\ref{allf}).  
Thus, from Newton's equations, all (nonzero) solutions $\alpha_+(t)$ are proportional to each other, and similarly all  (nonzero) $\alpha_-(t)$
are proportional to each other.  
The parameter choice in Eq. (\ref{abs}) assures that ${\alpha}_+(t_b)=\dot{\alpha}_+(t_b)=0$ for all configurations.   
Fixing, for example $\alpha_-(\uparrow\uparrow)(t_b)=0$, $a_3$ may be fixed as in  
Eq. \eqref{a3diff}, so that  $\dot{\alpha}_-(t_b)=0$, and therefore ${\alpha}_-(t_b)=\dot{\alpha}_-(t_b)=0$ as well for all configurations.  
Using Eq. \eqref{phase} to calculate the phases, and 
imposing $\Delta\phi=\gamma$, the remaining parameter (${a}_4$) is fixed as
\beq
\label{a4tilde}
{a}_4=C a_4^0,
\eeq
where  $a_4^0\equiv a_4(c_1=c_2=-1)$ is given in Eq. \eqref{a4diff}
and 
\beq
C=2\sqrt{\frac{-c_1}{(c_1-1)(c_2-1)}}.
\eeq
All coefficients in $\alpha_+(\uparrow\downarrow)$ are proportional to $a_4$,
so $\alpha_+(\uparrow\downarrow)$ is just scaled by the factor C with respect to the 
ones for $c_1=c_2=-1$ in the main text, and  $f_+(\uparrow\downarrow)$  is also scaled by the same factor according to 
Eq. (\ref{auxiliarya}).  
Comparing Eqs. (\ref{FaFb}) and (\ref{FaFb0}), and using $f_-(\ua\da)=0$,  we find that   
\beqa
\label{FaFbtilde}
\tilde{F}_a &=& -\frac{C}{c_1} F_a,
\nonumber\\
\tilde{F}_b &=& CF_b, 
\eeqa
in terms of the forces $F_a,\,F_b$ given in Eq. (\ref{controlparamdiff}) for $c_1=c_2=-1$. 
All these functions have odd symmetry with respect to the middle time $t_f/2$ so that there is no contribution to the phase from 
the time integral of $\tilde{f}$, see Eq. (\ref{hanm}).  

Finally let us analyze the limit of equal masses where  $c_1=c_2=c$ and $\mu=1$. In the main text, this implies $c_1=c_2=c=-1$ and $F_a(\mu\rightarrow 1)=F_b(\mu\rightarrow 1)=F$, in agreement with the physical constraint of using the same laser for both ions. However, when $c\neq 1$, 
$\tilde{F}_a(\mu\rightarrow 1)\neq \tilde{F}_b(\mu\rightarrow 1)$,  see  Eq. \eqref{FaFbtilde}. Physically this implies the use 
of two different lasers which is not possible in practice, so equal masses with $c\neq 1$ have to be treated 
separately, as specified in Sec. \ref{emi}.     
\section{Integral expression for the phase\label{AppendixC}}
For $\alpha_\pm(0)=\dot\alpha_\pm(0)=0$,  Eq. (\ref{auxiliarya}) may be solved as 
$\alpha_\pm(t)=\frac{1}{\Omega_\pm}\int_0^t dt' f_\pm(t') \sin[\Omega_\pm(t-t')]$, see Eq. (\ref{gene}). Thus the phase (\ref{phase}) can be also expressed by  double integrals of the form 
\beqa
\label{phaseforces}
\phi(t_f)&=&\sum_{\mu=\pm}\!\int_0^{t_f}\!\! dt'\!\!\int_0^{t'}\! dt'' f_\mu(t')f_\mu(t'')
\frac{\sin[\Omega_\mu(t'-t'')]}{2\hbar\Omega_\mu}\hspace{.45cm}
\nonumber\\
\!\!&=&\sum_{\mu=\pm}\!\int_0^{t_f}\!\!\!\!\!\int_0^{t_f}\!\! dt'dt'' f_\mu(t')f_\mu(t'')
\frac{\sin(\Omega_\mu|t'-t''|)}{4\hbar\Omega_\mu},
\nonumber\\
\eeqa
see also \cite{Garcia-Ripoll2005,Staanum2004,Vitanov2014}. 
\section{Worst case fidelity
\label{AppendixD}}
To simplify notation, 
let us denote the internal state configurations by a generic index $s={\ua\ua, \ua\da, \da\ua, \da\da}$. 
Assume an initial state $|\psi_m\ra(\sum_s c_s |s\ra)$, where $\sum |c_s|^2=1$ and the ``m'' here stands for ``motional''. 
The ideal output state, up to a global phase factor, 
is
\beq
|\psi_{id}\ra = \left(\sum_s c_s e^{i\phi(s)} |s\ra\right)|\psi_m\ra,
%
\eeq
where 
%
\beq
\phi(\ua\da)+\phi(\da\ua)-\phi(\ua\ua)-\phi(\da\da)=\pm \pi.
\eeq
The actual output state is generally entangled, 
\beq
|\psi_{ac}\ra = \sum_sc_s e^{i\phi'(s)} |s\ra |\psi_{ms}\ra, 
\eeq
with a different motional state $|\psi_{ms}\ra$ for each spin configuration, and actual phases 
$\phi'(s)$.  
First we can compute the total overlap 
\beq
\la\psi_{id}|\psi_{ac}\ra
=
\sum_s |c_s|^2 e^{i[\phi'(s)-\phi(s)]}
\la\psi_m|\psi_{ms}\ra. 
\eeq
Moreover, writing each motional overlap in the form 
$\la\psi_m|\psi_{ms}\ra=|\la\psi_m|\psi_{ms}\ra|e^{i\phi_{ms}}=\epsilon_s e^{i\phi_{ms}}$, we have
\beqa
\la\psi_{id}|\psi_{ac}\ra &=& \sum_s 
|c_s|^2 \epsilon_s e^{i \delta_s}=\Re+i\Im,
\eeqa
where  $\delta_s\equiv\phi'(s)-\phi(s)+\phi_{ms}$, and 
\beqa
\Re &=& \sum_s |c_s|^2 \epsilon_s \cos{\delta_s}, 
\nonumber\\
\Im &=&  \sum_s |c_s|^2 \epsilon_s \sin{\delta_s}.
\eeqa
%
 
%
%
%
%
%
The fidelity is 
\beqa
\mathcal{F} &=& |\Re+ i \Im|^2= \Re^2 + \Im^2\ge \Re^2
\nonumber\\
&=& (\sum_s |c_s|^2 \epsilon_s \cos{\delta_s})^2.
\eeqa
Assuming a ``good gate'', such that $|\delta_s|\ll 1$ for all $s$, 
%
%
%
%
then the fidelity is bounded from below by the worst possible case, 
\beq
\label{worstfidelity}
\mathcal{F} \ge {\rm Min}[(\epsilon_s \cos{\delta_s})^2].
\eeq
%

%
%
%
%
\section{Spread of the position of one ion in the ground state of the two ions
\label{AppendixE}}
An approximate analytical wave function for the ground state of the two ions subjected to the Hamiltonian (\ref{zeroth}),  
is given by multiplying the 
ground states of the two normal modes, see Eq. \eqref{psistat},
\beq
\psi_{NM}=\left(\frac{\Omega_+\Omega_-}{\pi^2\hbar^2}\right)^{1/4}e^{-\frac{1}{2\hbar}(\Omega_+{\sf x}_+^2+\Omega_-{\sf x}_-^2)}.
\eeq
In laboratory coordinates, and  for the specific case of equal mass ions, the normalized ground state is 
\begin{widetext}
\beq
\psi(x_1,x_2) = \left(\frac{m\sqrt{3}\omega^2}{\pi^2\hbar^2}\right)^{1/4}e^{-\frac{m\omega}{4\hbar}\left[ (1+\sqrt{3})(x_1+\frac{x_0}{2})^2 +(1+\sqrt{3})(x_2-\frac{x_0}{2})^2+2(1-\sqrt{3})(x_1+\frac{x_0}{2})(x_2-\frac{x_0}{2}) \right]}.
\eeq
\end{widetext}
The expectation values of $x_1$ and $x_1^2$ are calculated as
\beqa
\langle x_1 \rangle &=& \int \limits_{-\infty}^{\infty}\int \limits_{-\infty}^{\infty} dx_1dx_2 x_1 \psi^2 (x_1,x_2)=-\frac{x_0}{2},\nonumber\\
\langle x_1^2 \rangle &=& \!\!\!\! \int \limits_{-\infty}^{\infty}\int \limits_{-\infty}^{\infty} dx_1dx_2 x_1^2 \psi^2 (x_1,x_2)=\frac{ x_0^2}{4}+\frac{(3+\sqrt{3})\hbar}{12m\omega} ,\nonumber\\
\eeqa
so that the wave packet width for ion 1 is
\beq
\Delta x_1=\sqrt{\langle x_1^2 \rangle-\langle x_1 \rangle^2}=\frac{1}{2}\sqrt{1+\frac{1}{\sqrt{3}}}\sqrt{\frac{\hbar}{m\omega}}.
\eeq
\section{Alternative inversion protocols\label{AppendixF}}
In the inversion protocol used in Sec. \ref{V} for different masses, we have set 
$\alpha_-(\uparrow\downarrow;t)=0$ so this mode does not produce any excitation or any phase. 
It is also possible to have the other mode at rest, $\alpha_+(\uparrow\downarrow;t)=0$, by plugging
\beqa
\label{alternativeansatz}
\alpha_+(\uparrow\downarrow) &=& 0,
\nonumber\\
\alpha_-(\uparrow\downarrow) &=& b_0+\sum_{n=1}^4 b_n \cos\left[\frac{(2n-1)\pi t}{t_f}\right].
\eeqa
When choosing the alternative ansatz in Eq. \eqref{alternativeansatz}, the critical times (\ref{criticaltimes}) are exactly the same.  
%
%
Alternatively, we may cancel one of the normal modes in the parallel configuration substituting the inversion scheme in Eq. \eqref{solscheme} by 
\beq
\alpha_\pm(\uparrow\uparrow) \dashrightarrow f_\pm (\uparrow\uparrow) \dashrightarrow F_a, F_b \dashrightarrow f_\pm(\uparrow\downarrow) \dashrightarrow \alpha_\pm(\uparrow\downarrow).
\eeq
This type of freedom may be useful to further minimize the forces. 

\begin{thebibliography}{99}
%
\bibitem{Sorensen1999} A. S\o rensen, and K. M\o lmer, Phys. Rev. Lett. \textbf{82}, 1971 (1999).
%
\bibitem{Solano1999} E. Solano, R. L. de Matos Filho, and N. Zagury,  Phys. Rev. A 59, R2539 (1999).
%
\bibitem{Milburn2000}G. J. Milburn, S. Schneider and D. F. V. James, Fortschr. Phys. \textbf{48}, 801 (2000). 
%
\bibitem{Sorensen2000} A. S\o rensen and K. M\o lmer, Phys. Rev. A \textbf{62}, 022311 (2000). 
%
\bibitem{Sackett} C. A. Sackett et al., Nature \textbf{404}, 256 (2000).
%
\bibitem{Leibfried2003} D. Leibfried, B. DeMarco, V. Meyer, D. Lucas, M. Barrett, J. Britton, W. M. Itano, B. Jelenkovi\'c, C. Langer, T. Rosenband, and D. J. Wineland, Nature \textbf{422}, 412 (2003).
%
\bibitem{Garcia-Ripoll2003} J. J. Garc\' ia-Ripoll, P. Zoller, and J. I. Cirac, Phys. Rev. Lett. \textbf{91}, 187902 (2003).
%
\bibitem{Garcia-Ripoll2005} J. J. Garc\' ia-Ripoll, P. Zoller, and J. I. Cirac, Phys. Rev. A \textbf{71}, 062309 (2005).
%
%
\bibitem{Staanum2004} P. Staanum, M. Drewsen, and K. M\o lmer, Phys. Rev. A \textbf{70}, 052327 (2004).
%

\bibitem{Steane2014} A. M. Steane, G. Imreh, J. P. Home, and D. Leibfried, New J. Phys. \textbf{16}, 053049 (2014).
%

\bibitem{Vitanov2014} L. S. Simeonov, P. A. Ivanov, and N. V. Vitanov, Phys. Rev. A \textbf{89}, 012304 (2014).
%
\bibitem{Bentley2013}  C. D. B. Bentley, A. R. R. Carvalho, D.  Kielpinski, and J. J. Hope, New J. Phys. \textbf{15}, 043006 (2013).
%
\bibitem{Taylor2016} R. L. Taylor, C. D. B. Bentley, J. S. Pedernales, L. Lamata, E. Solano, A. R. R. Carvalho, and J. J. Hope, ArXiv1601.00359 (2016).
%
\bibitem{Duan2004} L.-M. Duan, Phys. Rev. Lett. \textbf{93}, 100502 (2004).
%
\bibitem{Lewis1969} H. R. Lewis and W. B. Riesenfeld, J. Math. Phys. \textbf{10}, 1458 (1969).
%
\bibitem{Lee2005} P. J. Lee, K.-A. Brickman, L. Deslauriers, P. C. Haljan, L.-M. Duan, and C. Monroe, J. Opt. B: Quantum Semiclass. Opt. \textbf{7}, S371 (2005).
%

 %
\bibitem{Tan2015} T. R. Tan, J. P. Gaebler, Y. Lin, Y. Wan, R. Bowler, D. Leibfried, and D. J. Wineland, Nature \textbf{528}, 380-383 (2015).
%
\bibitem{Sasura2003} M. Sasura and A. M. Steane, Phys. Rev. A \textbf{67}, 062318 (2003). 
%
\bibitem{Chen2010} X. Chen, A. Ruschhaupt, S. Schmidt, A.  del Campo, D. Gu\'ery-Odelin, and J. G. Muga, Phys. Rev. Lett. \textbf{104}, 063002 (2010). 
%
\bibitem{Torrontegui2013} E. Torrontegui, S. Ib\' a\~nez, S. Mart\'\i nez-Garaot, M. Modugno, A. del Campo, D. Gu\'ery-Odelin, A. Ruschhaupt, X. Chen, and J. G. Muga, Adv. At. Mol. Opt. Phys. \textbf{62}, 117 (2013).
%
%
\bibitem{Torrontegui2011} E. Torrontegui, S. Ib\' a\~nez, X. Chen, A. Ruschhaupt, D. Gu\'{e}ry-Odelin, and J. G. Muga, Phys. Rev. A \textbf{83}, 013415 (2011).
%
\bibitem{Palmero2014} M. Palmero, R. Bowler, J. P. Gaebler, D. Leibfried, and J. G. Muga, Phys. Rev. A \textbf{90}, 053408 (2014). 
%
\bibitem{Leibfried2007} D. Leibfried, E. Knill, C. Ospelkaus, and D. J. Wineland, Phys. Rev. A \textbf{76}, 032324 (2007).
%
\bibitem{Kosloff1988} R. Kosloff, J. Phys. Chem. \textbf{92}, 2087 (1988).
%
%
\bibitem{Kosloff1986} R. Kosloff and H. Tal-Ezer, Chem. Phys. Lett. \textbf{127}, 223 (1986). 

\bibitem{Andreas} A. Ruschhaupt, X. Chen, D. Alonso, and J. G. Muga, New J. Phys. \textbf{14}, 093040 (2012).
%
\bibitem{Lu} X.-J. Lu, X. Chen, A. Ruschhaupt, D. Alonso, S. Gu\'erin, and J. G. Muga, Phys. Rev. A \textbf{88}, 033406 (2013).
%
\bibitem{Lu2} X.-J. Lu, J. G. Muga, X. Chen, U. G. Poschinger, F. Schmidt-Kaler, and A. Ruschhaupt,
Phys. Rev.  A \textbf{89}, 063414 (2014).
%
\bibitem{Daems} D. Daems, A. Ruschhaupt, D. Sugny, S. Gu\'erin, Phys. Rev. Lett. \textbf{111}, 050404 (2013).
 



%
\end{thebibliography}
\end{document}